\documentclass[journal=jctc,manuscript=article, layout = twocolumn,]{achemso}
\usepackage[version=3]{mhchem}

 \SectionNumbersOn
\usepackage[english]{babel}
\usepackage{microtype}
\usepackage{indentfirst}
\usepackage{caption}
\usepackage{graphicx}
\usepackage{booktabs}
\usepackage{tabularx}
\usepackage{multirow}
\usepackage[caption=false]{subfig}
\usepackage{braket}
\usepackage{amsmath}
\usepackage{array}
\usepackage{dcolumn}
\usepackage{bm}
\usepackage{setspace}
\usepackage{float}
\usepackage{ulem}
\usepackage{color}
\usepackage{rotating}
\usepackage[varg]{txfonts}
\usepackage[font=footnotesize,labelfont=bf]{caption}
\usepackage[small]{titlesec}

\newcommand\eqt{\hspace{0.17em}{=}\hspace{0.17em}}

\newcommand\pt{\hspace{0.17em}{+}\hspace{0.17em}}
\newcommand\mt{\hspace{0.17em}{-}\hspace{0.17em}}
\newcommand\sd{\hspace{0.05em}}
\newcommand\kt{\hspace{0.17em}{<}\hspace{0.17em}}

\newcommand\get{\hspace{0.17em}{\ge}\hspace{0.17em}}

\newcommand\intext{\hspace{0.15em}{\in}\hspace{0.15em}}
 
  \newcommand\timest{\hspace{0.12em}{\times}\hspace{0.12em}}

\newcommand\br{\mathbf{r}}
\newcommand\bG{\mathbf{G}}
\newcommand\rc {r_\text{c}}
\newcommand\deft{\hspace{0.17em}{=:}\hspace{0.17em}}
\newcommand{\GW}{\textit{GW} }

\newcommand{\mrc}{m_{\rc }}

\newcommand{\bL}{\mathbf{L}}
\newcommand{\bLT}{\mathbf{L}^\text{T}}

\newcommand{\bchitil}{\boldsymbol{\tilde{\chi}}^0}
\newcommand{\Nminimax}{N}
\newcommand{\bV}{\mathbf{V}}
\newcommand{\chidetails}{\mathbf{M}^{-1}\boldsymbol{\chi}^0(i\omega)\mathbf{M}^{-1}}

\definecolor{darkgreen}{rgb}{0.0,0.6,0.0}

\newcommand{\JW}[1]{{\color{black} #1}}

\title{Low-scaling \textit{GW} with benchmark accuracy and application to phosphorene nanosheets}

  \author{Jan Wilhelm}\email{jan.wilhelm@physik.uni-regensburg.de}\affiliation{Institute of Theoretical Physics, University of Regensburg, D-93053 Regensburg, Germany}
  \author{Patrick Seewald}\affiliation{Department of Chemistry, University of Zurich, CH-8057 Zurich, Switzerland}
   \author{Dorothea Golze}\affiliation{Department of Applied Physics, Aalto University,  FI-00076 Aalto, Finland}

\let\oldmaketitle\maketitle
\let\maketitle\relax

\begin{document}
\linespread{1.1}
\fontsize{10}{12}\selectfont
\bibliographystyle{plainnat}
\normalem

 \twocolumn[
  \begin{@twocolumnfalse}
    \oldmaketitle
    \begin{abstract}
\fontsize{10}{12}\selectfont
$GW$ is an accurate method for computing electron addition and removal energies of molecules and solids.
In a conventional $GW$ implementation, however, its computational cost is $O(N^4)$ in the system
size $N$, which prohibits its application to many systems of interest.
We present a low-scaling $GW$  algorithm with notably improved accuracy compared to our previous algorithm {[J.~Phys.~Chem.~Lett.~\textbf{2018}, 9, 306\,--\,312]}. This is demonstrated for frontier orbitals using the $GW100$ benchmark set, for which our algorithm yields a mean absolute deviation of only 6~meV with respect to canonical implementations. We show that also excitations of deep valence, semi-core and unbound states match conventional schemes within 0.1~eV.
The high accuracy is achieved by using mini\-max grids with 30 grid points and the resolution of the identity with the truncated Coulomb metric.
We apply the low-scaling $GW$ algorithm with improved accuracy to phosphorene nanosheets of increasing size.
We find that their fundamental gap is strongly size-dependent varying from 4.0~eV (1.8\,nm\,$\times$\,1.3\,nm, 88 atoms) to 2.4~eV (6.9\,nm\,$\times$\,4.8\,nm, 990 atoms) at the ev$GW_0$@PBE level.
\vspace{0.5em}
  \end{abstract}
  \end{@twocolumnfalse}
  ]

\section{Introduction}
The $GW$ approximation\cite{Hedin} to many-body perturbation theory has become the method of choice for the calculation of photoemission spectra of materials and more recently also of molecules.\cite{GWreview2019,Reining2017} The extension of $GW$ to the Bethe-Salpeter equation\cite{Salpeter1951} has been extensively applied for the accurate computation of absorption spectra in materials science\cite{RevModPhys.74.601}
and chemistry\cite{Blase2018,Blase2020} \JW{and lately also to ground- and excited-state geometry optimizations.\cite{Berger2021, Caylak2021}} Recent $GW$ trends include the application to deep core excitations,\cite{Aoki2018,Dorotheacontroudef,coreJPCL,Keller2020,zhu2020all} comprehensive benchmarking\cite{GW100,GW100VASP,GW100pwloc,WestGW100,cubicGW,RANGEL2020,Foerster2020a,sepDFGaoChelikowsky} and the development of computationally efficient schemes for large-scale calculations of systems with $\geq$\,1000 atoms.\cite{cubicGW,DelBen2019,ONGW} This work contributes to the last two points with focus on avoiding the loss of numerical accuracy with respect to canonical $GW$ implementations.\par 
The application of conventional $GW$ schemes is restricted to systems with a few hundred atoms,\cite{GWCP2K,Stuke2020} due to the $O(N^4)$ scaling with respect to system size $N$ and the large overall computational cost (prefactor).
Recent developments to make larger system sizes computationally tractable cover the range from massively parallel implementations over physically motivated approximations to novel numerical methods. Efficient parallelization schemes were developed for execution on more than 10,000 CPU cores\cite{cubicRPAcp2k,cubicGW,DelBen2019,Kim2019,Sangalli2019} and first algorithms have been already proposed for the new generation of heavily GPU-based (pre)exascale supercomputers.\cite{DelBen2020} An example for more physically motivated developments are $GW$ embedding schemes, where a small part of the system is calculated at the $GW$ level and the surrounding medium is treated at a lower level of theory.\cite{Blasepol,Li2016,Li2018} Numerical developments have proceeded in several directions, either reducing the computational prefactor or the scaling with respect to system size. \par
%
The prefactor has been reduced by avoiding the summation over unoccupied states by solving the Sternheimer equation.\cite{PhysRevB.81.115105,Umari2,Lambert2013,GallioldPRB,Galli,SCHLIPF2020} A different strategy to reduce the overall computational cost are low-rank approximations of the polarizability, which map the latter onto a smaller basis.\cite{Wilson2008,Wilson2009,Galli,DelBen2019} Others addressed the frequency integration\cite{Friedrich2019,robustconti,Bintrim2020} or explored real-space density fitting schemes.\cite{sepDFGaoChelikowsky} The size of the matrices can be also reduced by choosing an optimal basis set for the respective problem. Localized basis sets are generally smaller than traditional plane-wave basis sets and particularly suited for molecular systems. The implementation of $GW$ in quantum chemistry codes, which typically use localized basis set, is a rather recent development of the last decade.\cite{Blasealt,Ren2012,FEGW,molgwimpl,GWCP2K,cubicGW,Foerster2020a,Sun2020}
The efficient inclusion of periodic boundary conditions into algorithms with localized basis sets is still subject of on-going work. \cite{periodicGWCP2K,zhu2020all,ren2020,solvelectron,Iskakov2020}\par
Scaling reduction is a particularly promising approach when aiming at application to nanostructured systems, which require very large system sizes with 1000 atoms and more. Different approaches have been explored for the reduction of the scaling with respect to system size. A linear scaling algorithm was devised within the framework of stochastic $GW$.\cite{ONGW} While the stochastic schemes have been successfully applied to silicon,\cite{ONGW} the application to molecules seems to be more challenging.\cite{Vlcek2017} Several cubic-scaling algorithms were developed,\cite{Scaling,liu2016cubic,cubicGW,Duchemin2019,Foerster2020a,Kim2020} which are based on or at least inspired by the space-time method proposed by Rojas, Godby and Needs in 1995.\cite{AC1} Variants of the space-time method have been implemented in a plane-wave/projector-augmented-wave (PAW) $GW$ code\cite{liu2016cubic} and also with localized basis sets using Gaussian\cite{cubicGW,Duchemin2019} and Slater-type orbitals.\cite{Foerster2020a}\par
In our recent work,\cite{cubicGW} we devised a low-scaling $GW$ algorithm in a Gaussian basis with formal $O(N^3)$ complexity, which has been optimized for massively parallel execution. Sparse linear algebra was exploited by using the resolution-of-the-identity (RI) approach with an overlap metric to refactor the four-center electron repulsion integrals. We showed that our algorithm scales effectively $O(N^2)$ and we applied it to quasi-one-dimensional systems (graphene nanoribbons) with more than 1700 atoms and 5700 electrons. An important property of low-scaling algorithms is the crossover point. The latter refers to the system size, where the low-scaling algorithm, which has usually a larger computational prefactor, becomes computationally more efficient than the canonical scheme. We demonstrated that the crossover point is already at around 150 atoms.\cite{cubicGW} \par
Another challenge for low-scaling $GW$ algorithms is reaching high numerical accuracy.\cite{Foerster2020a,Vlcek2017}
The $GW100$ benchmark\cite{GW100} has set the accuracy standards for molecules. Using identical basis sets, it was demonstrated that it is possible to match $GW$ excitations of the highest (HOMO) and lowest occupied molecular orbital (LUMO) within $<10$\,meV\cite{GW100} between two $GW$ implementations\cite{Ren2012,FEGW} based on numerically very different techniques.
For our previous low-scaling algorithm,\cite{cubicGW} we found that the $GW100$ mean absolute deviation (MAD) with respect to the canonical reference implementation in FHI-aims\cite{Ren2012} is 35~meV for ionization potentials and 27~meV for electron affinities. In addition, a couple of outliers with deviations in the range of 200~meV were observed, see \mbox{Ref.~\citenum{cubicGW}} (supporting information) and Ref.~\citenum{GWreview2019} for a comparison of the accuracy of different implementations.\par
The goal of this work is to increase the accuracy of the low-scaling $GW$ algorithm towards benchmark accuracy, i.e., MADs of less than 10~meV for the $GW100$ test, while retaining high computational efficiency. Furthermore, we aim to increase the reliability of our algorithm by reducing the number of outliers. High accuracy is achieved by a two-fold approach. The first is an increase and dedicated optimization of the mini\-max time and frequency grids, which can be directly transferred to other implementations of the space-time method. Second, we replace the overlap RI metric by the truncated Coulomb metric (RI-tC). In this work, the RI-tC approach is explored in the context of $GW$ for the first time.\par  
The remainder of this article is organized as follows: 
In Section~\ref{sec:GWspacetimebasic}, the $GW$ space-time method\cite{AC1,AC2} is introduced in a real-space grid formulation for non-periodic systems. 
The RI-tC approach is discussed in Section~\ref{sec:tC}.
Combining both, the $GW$ space-time method and the RI-tC within a Gaussian basis, we arrive at our low-scaling $GW$ algorithm (Section~\ref{GWGaussianRItrunc}).
Implementation details and computational details are given in Sections~\ref{sec:impldetails} and~\ref{sec:compdetails}, respectively.
Convergence tests of the mini\-max grid and the RI-tC are reported in Section~\ref{sec:accuracy}, including benchmark studies for the $GW$100 test set. We demonstrate that our low-scaling algorithm is not only accurate for frontier orbitals, but also for semi-core and unbound states by comparing to highly accurate contour-deformation results from the FHI-aims code\cite{Dorotheacontroudef} in Section~\ref{sec:alllevels}.
We apply our new low-scaling scheme to compute fundamental gaps of phosphorene nanosheets, which show potential as novel two-dimensional semiconductors, in Section~\ref{sec:appl}. \JW{Finally, we discuss the computational efficiency of our implementation in Section~\ref{sec:computeff} and} draw conclusions in Section~\ref{sec:conclusion}.

\section{\GW space-time method in a real-space formulation}\label{sec:GWspacetimebasic}
The $GW$ space-time method was proposed by Rojas, Godby, and Needs in 1995,\cite{AC1} enabling the computation of $GW$ quasiparticle (QP) energies at $O(N^3)$ complexity.
The approach by Rojas \textit{et al.} targets the application to solids employing a real-space grid in combination with a plane-wave basis.
%
Fast Fourier transforms are used to change the representation from the real-space grid to plane waves introducing a large computational prefactor. To keep the computational cost tractable, the original space-time approach is typically used together with soft pseudopotentials.\cite{AC2} This implies that deep valence or semi-core states are not included in the calculation of the density response functions making the application to materials with, e.g., localized $d$ electrons difficult.\par
%
%
The $GW$ space-time method was adapted to the PAW methodology by Liu \textit{et al.} in 2016,\cite{liu2016cubic} enabling the inclusion of more localized states in the density response function. 
The PAW implementation in VASP allows the efficient treatment of molecules\cite{GW100VASP} and large supercells\cite{liu2016cubic} with high accuracy.
However, the large computational prefactor due to the fast Fourier transforms between real and reciprocal space remains, similarly as in the original method.\cite{AC1}\par 
%

%
%
%
%
Fast Fourier transforms can be circumvented by replacing the real-space grid and the plane-waves basis by a localized basis, which was first explored in our work from 2018\cite{cubicGW} and very recently also by F\"orster and Visscher.\cite{Foerster2020a} In our work from 2018, we used a Gaussian basis in combination with a local metric (overlap) for the RI refactorization of the four-center Coulomb integrals. 
The low-scaling $GW$ algorithm developed by F\"orster and Visscher\cite{Foerster2020a} employs Slater-type functions. Unlike in our approach, sparsity is introduced by a local RI scheme (pair-atomic density fitting) instead of a local metric. We elaborate on the difference, advantages and disadvantages in Section~\ref{sec:globalvslocal}. \par
%
%

%
An alternative reformulation of the space-time method was proposed by Duchemin and Blase,\cite{Duchemin2019} combining a real-space grid with a Gaussian basis instead of plane waves. 
The real-space grid is specifically optimized for the respective molecule by the separable resolution of the identity. In \mbox{Ref.~\citenum{Duchemin2019}}, the described approach was only applied to the random phase approximation (RPA), but the extension to $GW$ is straightforward.\par
%
%

The aforementioned space-time algorithms\cite{AC1,AC2,liu2016cubic,cubicGW,Foerster2020a,Duchemin2019} differ in the choice of the basis and the associated numerical techniques. However, the time and frequency treatment is identical. To introduce the basic equations, we start with a generic reformulation of the $GW$ space-time algorithm for non-periodic systems projecting all quantities on real-space grids. 
%
Note that these generic expressions differ from the original work by Rojas \textit{et al.}, where only some quantities are computed on real-space grids, e.g., the polarizability, and others, e.g., the dielectric function, in a plane-wave basis. 
%
%
In Section~\ref{sec:tC} and~\ref{GWGaussianRItrunc}, we will project these generic expressions into a Gaussian basis.
We start from a self-consistent Kohn-Sham density functional theory (KS-DFT) calculation.
The total energy of a many-electron
system in KS-DFT is obtained by solving the eigenvalue
problem
\begin{align}
    \Big(h^0(\br)+v_\text{xc}(\br)\Big)\,\psi_n(\br) = \varepsilon_n\,\psi_n(\br)\,.\label{e1}
\end{align}
$h_0(\br)$ contains the external and the Hartree potential as well as the kinetic energy, while the   exchange-correlation potential~$v^\text{xc}(\br)$ accounts for electron-electron interaction beyond the Hartree interaction. 
In the $GW$ space-time method, we use molecular orbitals (MOs) $\psi_n(\br)$ and eigenvalues $\varepsilon_n$  for computing the single-particle Green's function in imaginary time as
\begin{align}
\begin{split}
G(\br,\br',i\tau)= \left\{ 
\begin{array}{ll}
  i\sum\limits_i^\text{occ} \psi_i(\br)\psi_i(\br')\exp(\varepsilon_i\tau)\,, &  \tau >0\,,
\\[0.5em]
 - i\sum\limits_a^\text{virt} \psi_a(\br)\psi_a(\br')\exp(\varepsilon_a\tau)\,, &  \tau <0\,.
\end{array}
\right.
\end{split}
\label{Greensfrrprime}
\end{align}
The irreducible polarizability is computed as
\begin{align}
    \chi^0(\br,\br',i\tau) = -iG(\br,\br',i\tau)G(\br,\br',-i\tau)\,.\label{e3}
\end{align}
We proceed by a Fourier transform to imaginary time to evaluate the dielectric function and its inverse as
\begin{align}
    \epsilon(\br,\br',i\omega) &= \delta(\br,\br') - 
    \int d\br''\,v(\br,\br'')\,\chi^0(\br'',\br',i\omega)\,,\label{e4}
    \\[1em]
     \epsilon^{-1}(\br,\br',i\omega) &= \delta(\br,\br') + 
    \int d\br''\,v(\br,\br'')\,\chi^0(\br'',\br',i\omega)+\ldots \label{e5}
\end{align}
with the bare Coulomb interaction $v(\br,\br')\eqt1/|\br\mt\br'|$ 
and using  $(1\mt x)^{-1}\eqt 1\pt x\pt x^2\pt\ldots\,$ for $|x|\kt 1$ in Eq.~\eqref{e5}.
The screened Coulomb interaction~$W$ is then given by
\begin{align}
    W(\br,\br',i\omega) &= \int d\br''\,\epsilon^{-1}(\br,\br'',i\omega)\,v(\br'',\br')\,.\label{e6}
\end{align}
Note that Eqs.~\eqref{e4}\,--\,\eqref{e6} are not implemented in real-space in any of the discussed space-time algorithms  because the computational cost of Eqs.~\eqref{e4}\,--\,\eqref{e6} quickly grows as~$N_\text{grid}^3$ with the number of real-space grid points~$N_\text{grid}$, which prohibits the application to large systems.
In the original space-time method,\cite{AC1} Eqs.~\eqref{e4} and~\eqref{e6} are formulated in plane-waves with a diagonal Coulomb operator~$V_{\bG\bG'}\eqt\delta_{\bG\bG'}/|\bG|^2$ such that the scaling of Eq.~\eqref{e4} and~\eqref{e6} is reduced to $O(N^2)$.
%

%

%
We continue the algorithm by a Fourier transform of~$W(i\omega)$ from Eq.~\eqref{e6} to imaginary time to evaluate the self-energy as
  \begin{align}
    \Sigma(\br,\br',i\tau) = iG(\br,\br',i\tau)W(\br,\br',i\tau)\,.\label{sigmarrprimetau}
\end{align}
After Fourier transforming $\Sigma(i\tau)$ to imaginary frequency~$i\omega$, we use analytic continuation to obtain $\Sigma(\omega)$ such that the $G_0W_0$ QP energies can be evaluated as
\begin{align}
\varepsilon_n^{G_0W_0} = \varepsilon_n
+ \text{Re}\;\Sigma_n(\varepsilon_n^{G_0W_0}) -v^\text{xc}_n 
\label{qpeq}
\end{align}
where $v^\text{xc}_n$ and $\Sigma_n(\varepsilon)$ are $(n,n)$-diagonal  matrix elements in the MO basis~$\psi_n$ of the respective quantities, 
\begin{align}
    \Sigma_n(\varepsilon) &= \int d\br\, d\br'\,
    \psi_n(\br)\,\Sigma(\br,\br',\varepsilon)\,\psi_n(\br')\,,
    \\[0.5em]
    v^\text{xc}_n &= \int d\br\;
     \psi_n(\br)\,v^\text{xc}(\br)\,\psi_n(\br)\,.
\end{align}
In this work, we will also use eigenvalue-selfconsistent $GW_0$ (ev$GW_0$) where~$\varepsilon_n^{G_0W_0}$ are used to recompute $G(i\tau)$ from Eq.~\eqref{Greensfrrprime}.
$\Sigma(i\tau)$ follows from Eq.~\eqref{sigmarrprimetau} using $W(i\tau)$ from $G_0W_0$.
The QP energy is recomputed from Eq.~\eqref{qpeq}.
In ev$GW_0$, this cycle is repeated until the QP energies are converged.

\begin{figure}[t]
\centering
\includegraphics[width=0.48\textwidth]{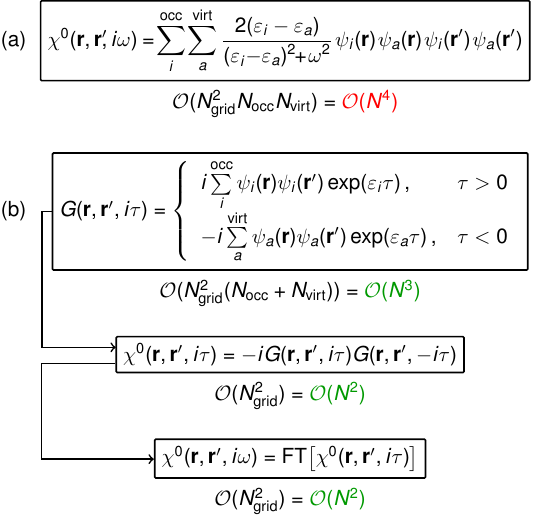}
\caption{
Computation of the irreducible polarizability (a) in an ordinary~$O(N^4)$ implementation\cite{GWreview2019} and (b) in the $GW$ space-time method.\cite{AC1}
In most $GW$ algorithms, this step dominates the computational cost of the whole $GW$ calculation.
In (a), the computational cost increases as $N^4$ with the system size~$N$ since the following quantities each increase linearly with~$N$: the number of real-space grid points~$N_\text{grid}$, the number of occupied molecular orbitals~$N_\text{occ}$ and the number of virtual molecular orbitals~$N_\text{virt}$. In (b), we repeat Eqs.~\eqref{Greensfrrprime} and~\eqref{e3}. Calculating the irreducible polarizability in imaginary frequency is reduced to $O(N^3)$ scaling.
}
\label{scalingsketch}
\end{figure}
The scaling of the different steps is summarized Fig.~\ref{scalingsketch}. In a canonical implementation, the evaluation of the polarizability is the computational bottleneck and scales with $O(N^4)$, see Fig.~\ref{scalingsketch}\,(a). The space-time method decouples the summation over occupied and virtual states in the polarizability by expressing $G$ in the time instead of the frequency domain, see Eq.~\eqref{Greensfrrprime}. This reduces the scaling to at most cubic, as shown in Fig.~\ref{scalingsketch}\,(b).
%

\section{Resolution of the identity (RI) using the truncated Coulomb metric}\label{sec:tC}

\subsection{RI for four-center Coulomb integrals}
Before reformulating the $GW$ space-time method from Section~\ref{sec:GWspacetimebasic} in a Gaussian basis, we focus on four-center Coulomb integrals (4c-CIs) that are of central importance in $GW$ calculations with localized basis sets. 
These 4c-CIs, in Mulliken notation, are defined as 
\begin{align}
(\mu\nu|\lambda\sigma) := \int  d\br\sd d\br'\sd  {\phi_\mu(\mathbf{r}')\phi_\nu(\mathbf{r}')\phi_\lambda(\mathbf{r})\phi_\sigma(\mathbf{r})}\,\frac{1}{|\mathbf{r}-\mathbf{r}'|}\label{4cERIdef}
\end{align}
where  $\phi_\mu, \phi_\nu,\phi_\lambda$ and~$\phi_\sigma$ are atomic-orbital (AO) Gaussian basis functions.
Using the RI\cite{RI3,RIXavier,RPAFurche} approximation with a pre-defined metric 
\begin{align}
m{:}\;\,\mathbb{R}^3\timest\,\mathbb{R}^3\rightarrow[0,\infty)\label{e12}\,,
\end{align}
the 4c-CIs are factorized into two- and three-center integrals\cite{RI3}
\begin{align}
(\mu\nu|\lambda\sigma)_\text{RI} = \sum_{PQRS} (\mu\nu |P)_mM^{-1}_{PQ}V_{QR}M^{-1}_{RS}(S|\lambda\sigma)_m\,.\label{ERI4c}
\end{align}
$P,Q, R$ and $S$ refer to indices of  auxiliary RI Gaussian basis functions.
 $\mathbf{M}$ denotes the representation of the metric~$m$ in the auxiliary RI basis~$\{\varphi_P\}$,
\begin{align}
M_{PQ} = \int  d\br\, d\br'\sd\varphi_P(\mathbf{r})\,m(\mathbf{r},\mathbf{r}')\,\varphi_Q(\mathbf{r}')\label{Mmatrix}\,.
 \end{align} 
 The three-center integrals $(\mu\nu| P)_m$ are given by
\begin{align}
(\mu\nu| P)_m\equiv (P|\mu\nu)_m = \int  d\br\, d\br'\sd {\phi_\mu(\mathbf{r})\sd\phi_\nu(\mathbf{r})\,m(\mathbf{r},\mathbf{r}')\,\varphi_P(\mathbf{r}')}\,.
\label{3c}
\end{align}%

The bare Coulomb interaction~$1/|\br\mt\br'|$  from Eq.~\eqref{4cERIdef} is contained in the Coulomb matrix element~$V_{QR}$ in Eq.~\eqref{ERI4c} which is given by
 \begin{align}
V_{PQ} = \int  d\br\, d\br'\sd\varphi_P(\mathbf{r})\,\frac{1}{|\mathbf{r}-\mathbf{r}'|}\,\varphi_Q(\mathbf{r}')\,.\label{Vmatrix}
 \end{align} 
We compute the two-center integrals~\eqref{Mmatrix} and \eqref{Vmatrix} with a solid-harmonic-based \JW{analytical} integration scheme\cite{DorotheaIntegrale} and the three-center integrals from Eq.\,\eqref{3c} with the \JW{analytical} Obara-Saika recurrence scheme.\cite{OS} \JW{Both schemes are applicable to general interaction potentials $g(|\br_1{-}\br_2|)$,\cite{Ahlrichs2006,DorotheaIntegrale} which includes the overlap, Coulomb and  truncated Coulomb potential discussed in Section~\ref{sec:truncatedcoulomb}. The calculation of the integrals starts in both schemes from integrals over primitive $s$-functions. The analytical expressions for the $s$-type integrals are given in Ref.~\citenum{Ahlrichs2006} for overlap and Coulomb potential and in Ref.~\citenum{Guidonthesis} for the truncated Coulomb potential. 
Prescreening of the two and three-center integrals is applied dependent on the metric. For the truncated Coulomb metric, the integrals are screened based on the exponents, the distance between the Gaussians centers and the truncation radius. In addition, computed three-center integrals, which are sufficiently close to zero, are filtered out before calculating the $GW$ quantities.  
}

\subsection{Truncated Coulomb metric as convenient choice in low-scaling methods}
\label{sec:truncatedcoulomb}
\begin{figure}[t]
\centering
\includegraphics[width=0.48\textwidth]{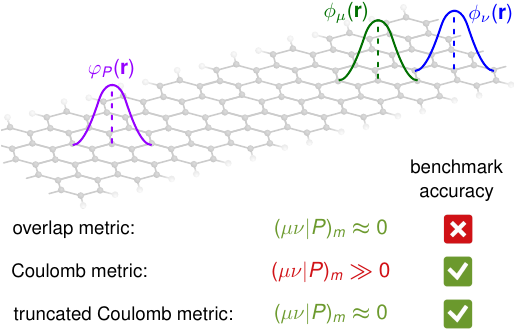}
\caption{
Sketch of three Gaussian basis functions, where the AO basis functions~$\phi_\mu(\br)$ and~$\phi_\nu(\br)$ are close together, while the RI basis function~$\varphi_P(\br)$ is far away from $\phi_\mu(\br)$,~$\phi_\nu(\br)$.
In this case, the three-center integral $(\mu\nu|P)_m$ from Eq.~\eqref{3c} vanishes in the overlap metric, and in the truncated Coulomb metric, while in the Coulomb metric the three-center integrals $(\mu\nu|P)_m$ are non-vanishing.
High accuracy in electronic structure methods can only be achieved by the Coulomb metric and the truncated Coulomb metric.\cite{RI3}
}
\label{sketchsparsity}
\end{figure}

The first key ingredient to reduce the scaling is the decoupling of the occupied and virtual MOs in the polarizability by working in the time domain.
The second ingredient, when working in a localized basis set, is the choice of the RI metric.
In our previous implementation of low-scaling $GW$,\cite{cubicGW} we employed the overlap metric\cite{RI3}
\begin{align}
m_\text{O}(\br,\br')\eqt\delta(\br\mt\br')
\end{align}
for computing the integrals in Eq.~\eqref{Mmatrix} and~\eqref{3c}, where $\delta$ is the Dirac distribution.
The overlap metric is local in the sense that the RI basis functions~$\varphi_P$ do not overlap with AO basis function products $\phi_\mu\phi_\nu$ in Eq.~\eqref{3c} if there is enough distance between their centers. This leads to vanishing three-center overlap matrix elements $(\mu\nu| P)_{m}$ and increasing computational efficiency due to sparsity, as illustrated in Fig.~\ref{sketchsparsity}.
In contrast, the Coulomb metric 
\begin{align}
m_\text{C}(\br,\br')=\frac{1}{|\br\mt\br'|}
\label{eq:mcmetric}
\end{align}
couples RI basis functions~$\varphi_P$ and AO basis function pairs~$\phi_\mu\phi_\nu$ in Eq.~\eqref{3c} over effectively infinite distances due to the slow polynomial decay of $1/|\br\mt\br'|$ as illustrated in Fig.~\ref{sketchsparsity}.\cite{Luenser}
With the Coulomb metric, no sparsity can be gained hampering its usage in low-scaling $GW$ algorithms.
%
%
In canonical $O(N^4)$ algorithms, each AO product~$\phi_\mu\phi_\nu$ is transformed to the delocalized molecular orbital basis~$\{\psi_n\}$ loosing all sparsity anyway.\cite{MauroRI,GWCP2K,vladimir,Ren2012,Luenser,RIWeigend,BSERI,GWSOC,GWstrongfield,periodicGWCP2K,minireview,Hutter2020}
In such a conventional algorithm, where sparsity cannot be exploited, the Coulomb metric is the optimal choice because the RI factorization given in Eq.~\eqref{ERI4c} converges much quicker with respect to the RI basis set size.\cite{RI3} The Coulomb metric yields thus generally higher accuracy than the overlap metric.

%
%
%

%
In this work, we improve our previous low-scaling $GW$ implementation\cite{cubicGW} by replacing the overlap metric by the truncated Coulomb metric\cite{jung2005auxiliary,HGtruncCoulSOSMP2,ReineRI,Luenser}
\begin{align}
\mrc(\br,\br') =  \left\{ 
\begin{array}{cl}
   \displaystyle \frac{1}{|\br-\br'|} & \text{if } |\br-\br'|<\rc \,,
   \\[0.5em]
    0  & \text{else}\,,
\end{array}
\right.\label{tC}
\end{align}
where the Coulomb interaction is cut after a distance~$\rc $.
In the limit of a large cutoff radius~$\rc $, the truncated Coulomb metric turns into the Coulomb metric, $\underset{\rc \rightarrow\infty}{\lim}\,m_{\rc }(\br,\br') \eqt{1}/{|\br-\br'|}$.
For a small cutoff radius~$\rc $, calculations based on the truncated Coulomb metric  are equivalent to calculations based on the overlap metric.\cite{Luenser}
The truncated Coulomb metric combines the attractive features of the Coulomb metric and the overlap metric: high accuracy due to the near-sighted Coulomb operator and preservation of sparsity due to the locality of~$m_{\rc }(\br,\br')$.
\JW{Another approach for truncating the Coulomb operator is the use of complementary error functions as in standard range-separated hybrid functionals.\cite{Heyd2003,Dutoi2008}}
 \JW{The benefits of a local Coulomb metric} have already been exploited for low-scaling scaled-opposite spin MP2\cite{HGtruncCoulSOSMP2} and low-scaling RPA\cite{Luenser,parallellsRPA,lsbeyondRPA,lsRPAgradients,lsRPAselfconsistent} reporting similar accuracy as for the respective conventional-scaling schemes.
The RI factorization in Eq.~\eqref{ERI4c} is exact in the limit of a complete RI basis, independent of the chosen RI metric. Therefore, truncating the Coulomb operator with a finite~$\rc $ does not affect the accuracy of the $GW$ algorithm as long as the RI basis is sufficiently large. 
%

%
We note that in plane-wave implementations, RI with different metrics is not discussed.
The reason is that the Coulomb matrix, the truncated Coulomb matrix and the overlap matrix are diagonal in the plane-wave basis. 
As consequence, RI factorizations as in Eq.~\eqref{ERI4c}  are identical for the three different metrics when using plane wave basis functions. We added a more detailed explanation in the supporting information (SI) to facilitate the discussion between plane-wave and localized-basis-set communities.
%

%
%

\subsection{Global vs. local RI}
\label{sec:globalvslocal}
The sums over the RI basis functions in the RI factorization of the 4c Coulomb integrals in Eq.~\eqref{ERI4c}
can either run over the whole RI basis ("global RI") or only over a subset of the RI basis ("local RI").
In their recent work, F\"orster and Visscher\cite{Foerster2020a} combined the $GW$ space-time method with the pair-atomic RI (PARI) approach.\cite{JCCMerlot} PARI, also known as pair-atomic density fitting (PADF) or RI-LVL,\cite{aimsLRIHF} is a local RI approach, which employs the Coulomb metric. Locality is introduced by expanding each AO pair~$\phi_\mu\phi_\nu$, where $\phi_{\mu}$ is centered at atom $A$ and $\phi_{\nu}$ at atom $B$, only in the subset of RI basis functions with centers at $A$ and $B$.\par
The scaling with PARI is the same as with global RI, if a local RI metric (overlap, truncated Coulomb) is employed for the latter.
However, PARI reduces the computational prefactor dramatically compared to global RI since the number of three-center integrals is substantially smaller. For example, the computational cost of a $GW$ calculation on $\approx$~400 atoms with around 8000 AOs is $\approx$~4000 CPU hours with our low-scaling scheme using global RI with the overlap metric,\cite{cubicGW} but only $\approx$~200 CPU hours with the PARI implementation by F\"orster and Visscher.\cite{Foerster2020a}
%
However, reaching high accuracy in low-scaling PARI-$GW$ seems more challenging.\cite{Foerster2020a}
\par
%
%
%
The accuracy of local RI schemes can be improved by adding high-angular-momentum functions to the RI basis set and increasing its size.\cite{aimsLRIHF,ren2020} It has been recently shown for a local RI variant of a $GW$ implementation with conventional scaling that good accuracy can be obtained with carefully chosen RI basis sets.\cite{ren2020} However, local RI schemes tend to ill-conditioning problems\cite{DorotheaLRI} introduced by very large RI basis sets, which might limit the attainable accuracy to some extent. It should be generally easier to reach high accuracy with MADs $\leq$\,10~meV, which is the focus of this work, with global RI-tC rather than a local PARI-type approach.

%

\section{\GW space-time method in a Gaussian basis using RI with the truncated Coulomb metric}\label{GWGaussianRItrunc}

In the following, we present our low-scaling $GW$ algorithm, which is a variant of the space-time method introduced in Section~\ref{sec:GWspacetimebasic}, and rationalize where the RI factorization from Eq.~\eqref{ERI4c} enters the algorithm. 

%
%

\subsection{Low-scaling algorithm}\label{subsec:GWalgoGauss}
%
The MOs $\{\psi_n\}$ are expanded in Gaussian-type orbitals (GTOs) $\{\phi_\mu\}$
\begin{align}
\psi_n(\mathbf{r}) = \sum_\mu C_{n\mu }\phi_\mu(\mathbf{r}),\label{expGauss}
\end{align}
where $C_{n\mu }$ are the MO coefficients.
%
The single-particle Green's function $G(i\tau)$ given in Eq.~\ref{Greensfrrprime} is then projected in the GTO basis
\begin{align}
\begin{split}
G_{\mu\nu}(i\tau)= \left\{ 
\begin{array}{ll}
  i\sum\limits_n^\text{occ} C_{n\mu}C_{n\nu}\exp(\varepsilon_n\tau)\,, &  \tau >0\,,
\\[0.5em]
 - i\sum\limits_n^\text{virt} C_{n\mu}C_{n\nu}\exp(\varepsilon_n\tau)\,, &  \tau <0\,.
\end{array}
\right.
\end{split}
\label{Greensf}
\end{align}
%

Next, we use $G(\br,\br',i\tau)\eqt{\sum_{\mu\nu}}\,\phi_\mu(\br)\,G_{\mu\nu}(i\tau)\,\phi_\nu(\br')$ and Eq.~\eqref{e3}, $    \chi^0(\br,\br',i\tau) \eqt {-}\sd iG(\br,\br',i\tau)G(\br,\br',-i\tau)$, to obtain the irreducible polarizability
$\chi^0(i\tau)$ in the Gaussian auxiliary  basis~$\{\varphi_P\}$\cite{cubicRPAcp2k,RIWeigend,MauroRI,EMSL}
\begin{figure}[t!]
\centering
\includegraphics[width=0.48\textwidth]{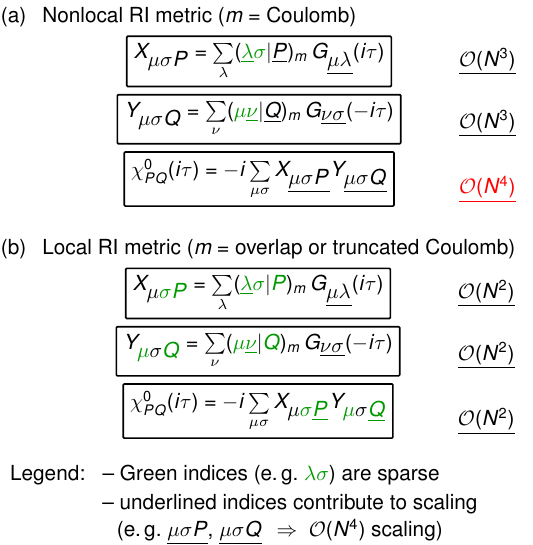}
\caption{
Scaling of the imaginary-time density response computation in a localized basis (Eq.~\eqref{NsquareRPA}) together with (a) a nonlocal and (b) a local RI metric.
$\chi^0_{PQ}(i\tau)$ is computed in two steps.\cite{cubicRPAcp2k} First the two tensors $X$ and $Y$ are computed followed by a tensor contraction.
Green color indicates sparse indices. The sparse index pair, e.g., $\lambda\sigma $ or sparse index triple, e.g., $\lambda\sigma P$ has together a scaling of $O(N)$.
Underlined indices contribute to the scaling. 
%
}

\label{fig:scalingmetric}
\end{figure}
\begin{align}
&{\chi}^0_{PQ}(i\tau)\coloneqq (\varphi_P|\chi^0|\varphi_Q)_m\nonumber
\\
&\coloneqq \int d\br_1\,d\br_2\,d\br_3\,d\br_4\,
\varphi_P(\br_1)\,m(\br_1,\br_2)\nonumber  
\\ &\hspace{6em}\times
\chi^0(\br_2,\br_3,i\tau)\,m(\br_3,\br_4)\,\varphi_Q(\br_4)\nonumber
\\[0.4em]
&=  -i\sum\limits_{\mu\sigma}\sum\limits_\lambda (\lambda\sigma| P)_m G_{\mu\lambda}(i\tau) \sum\limits_\nu (\mu\nu |Q)_m G_{\nu\sigma}(-i\tau)\,.\label{NsquareRPA}
\end{align}
The three-center integrals $(\mu\nu|P)_m$ are defined in Eq.~\eqref{3c} and originate from the RI factorization of the 4c-CIs given in Eq.~\eqref{ERI4c}. The expression for $\chi^0_{PQ}(i\tau)$ in Eq.~\eqref{NsquareRPA} is generic for any RI metric $m(\br,\br')$. The evaluation of $\chi^0_{PQ}(i\tau)$ is the computationally most expensive step and scales $O(N^4)$ with the conventional Coulomb metric (Eq.~\eqref{eq:mcmetric}). However, when employing a local RI metric, the three-center tensors $(\mu\nu|P)_m$ vanish unless the Gaussian functions $\phi_\mu$, $\phi_\nu$ and~$\varphi_P$ are centered on nearby atoms, which is illustrated in Fig.~\ref{sketchsparsity}. In this work, we use the local truncated Coulomb metric $m_{\rc }$ defined in Eq.~\eqref{tC}.
The computational complexity for the evaluation of $\chi^0_{PQ}(i\tau)$ reduces with a local metric to $O(N^2)$.\par 
A detailed analysis of the computational complexity of Eq.~\eqref{NsquareRPA} is shown in Fig.~\ref{fig:scalingmetric}. 
First, the multiplication of the three-center integrals with the Green's function~$G$ is computed, which yields the tensors $X$ and $Y$. The evaluation of $X$ and $Y$ scales cubically with a nonlocal metric, but only quadratically with a local metric.
%
The subsequent tensor contraction of $X$ and $Y$ is a step of $O(N^4)$ complexity with a nonlocal metric, which is reduced to $O(N^2)$ with the local variant.
The~$O(N^2)$ scaling behavior in  Fig.~\ref{fig:scalingmetric}\,(b) can be understood as follows: 
For computing a single matrix element~$\chi^0_{PQ}$ with a local metric, only a small $O(N^0)$-scaling number of~$\sigma$ indices (spatially close to $P$) and~$\mu$ indices (spatially close to $Q$) need to be taken into account.
Since the number of $PQ$-pairs increases as~$O(N^2)$, we end up with a final scaling of~$O(N^2)$ for the whole matrix~$\chi^0_{PQ}$.
\par
%
%
%
 %
%

%
We proceed by including the matrix elements $M_{PQ}$ from Eq.~\eqref{Mmatrix} in $\boldsymbol{\tilde{\chi}}^0(i\tau)$, 
\begin{align}
\boldsymbol{\tilde{\chi}}^0(i\tau)=\mathbf{M}^{-1}\boldsymbol{\chi}^0(i\tau)\mathbf{M}^{-1}\,.\label{chipseudoimtime}
\end{align} 
The polarizability~$\tilde{\boldsymbol{\chi}}^0(i\tau)$  is transformed to imaginary frequency via a cosine transform and the symmetric dielectric function~$\epsilon(i\omega)$ 
is computed by
\begin{align}
\boldsymbol{\epsilon}(i\omega)& =
 \mathbf{1}- \mathbf{L}^\text{T}\boldsymbol{\tilde{\chi}}^0(i\omega)\mathbf{L} \label{dielfunc} \,,
\end{align}
where $\mathbf{L}$ denotes the Cholesky decomposition of the Coulomb matrix~$\mathbf{V}$ from Eq.~\eqref{Vmatrix},
\begin{align}
 \mathbf{V}&= \mathbf{L}\mathbf{L}^\text{T}\,.\label{VLL}
\end{align}

The screened Coulomb interaction $W(i\omega)\eqt\epsilon^{-1}(i
\omega)V \eqt V + W^c(i\omega)$ is split into the bare Coulomb interaction
and the correlation contribution, and the latter 
is obtained as
\begin{align}
\mathbf{W}^\text{c}(i\omega) = \mathbf{L}\left[\boldsymbol\epsilon^{-1}(i\omega) -\mathbf{1}\right]\mathbf{L}^\text{T}\,,\label{Wc}
\end{align}
where the symmetric, positive definite $\boldsymbol{\epsilon}(i\omega)$ is inverted efficiently by Cholesky decomposition.
A cosine transform converts $W^\text{c}(i\omega)$ (Eq.~\eqref{Wc}) back to the imaginary time domain. 
Computing the quasiparticle energy for an orbital~$\psi_n$
  requires the   corresponding 
diagonal matrix element of the self-energy,
\begin{align}
\Sigma_n(i\tau) \eqt {\braket{\psi_n|\Sigma(i\tau)|\psi_n}}\deft\Sigma_n^x\pt\Sigma_n^\text{c}(i\tau)\,.
\end{align}
Its correlation part is obtained as
\begin{align}
\Sigma_n^\text{c}(i\tau) =i\sum_{\nu P}\sum_{\mu}G_{\mu\nu}(i\tau) (n\mu |P)_m\sum_Q\tilde{W}^\text{c}_{PQ}(i\tau)(Q|\nu n)_m\,,
\label{corrSEtaurewrite}
\end{align}
where 
$
\tilde{\mathbf{W}}^\text{c}(i\tau)\eqt\mathbf{M}^{-1}\mathbf{W}^\text{c}(i\tau)\mathbf{M}^{-1}
$,
and the exchange part is computed as
\begin{align}
\Sigma^\text{x}_n =  -
\sum_{\nu P} 
\sum_{\mu} D_{\mu\nu}(n\mu |P)_m \sum_Q \tilde{V}_{PQ} (Q|\nu n)_m\label{Sx}\,,
\end{align}
where
\begin{align}
D_{\mu\nu}&=\sum_n^\text{occ} C_{n\mu}C_{n\nu}\,,
\hspace{1em}
\tilde{\mathbf{V}}=\mathbf{M}^{-1}\mathbf{V}\mathbf{M}^{-1}\,.\label{DV}
\end{align}
%
%

%
In order to compute quasiparticle energies,
$\Sigma^\text{c}_n(i\tau)$ is transformed to imaginary frequencies by a sine and cosine transform.\cite{liu2016cubic} 
The self-energy is then evaluated on the real frequency axis $\Sigma^\text{c}_n(\varepsilon)$ by analytic continuation  of $\Sigma^\text{c}_n(i\omega)$.\cite{GW100,liu2016cubic,robustconti,GWCP2K}
The $G_0W_0$ energies~$\varepsilon_n^{G_0W_0}$ are obtained by solving the QP equation
\begin{align}
\varepsilon_n^{G_0W_0} = \varepsilon_n 
+\Sigma^\text{x}_n+\text{Re}\,\Sigma^\text{c}_n(\varepsilon_n^{G_0W_0})-v^\text{xc}_n 
\label{qpeq2}
\end{align}
iteratively for~$\varepsilon_n^{G_0W_0}$ via Newton-Raphson.\par
The calculation of the polarizability $\chi_{PQ}^0(i\omega)$ in Eq.~\eqref{NsquareRPA} remains also at $O(N^2)$ complexity the computational bottleneck, even for the largest systems studied in this work. The subsequent steps in  Eqs.~\eqref{chipseudoimtime}\,--\,\eqref{Wc} scale cubically, but have a much smaller computational prefactor. The calculation of the correlation self-energy from Eq.~\eqref{corrSEtaurewrite} scales as~$O(N^2)$ for every QP level~$n$ and is generally computationally less demanding than the calculation of ~$\boldsymbol{\chi}^0_{PQ}(i\tau)$.
%

\subsection{Tracing back four-center Coulomb integrals and RI factorizations}\label{subsec:discussionRIinGW}

While the full derivation of the algorithm presented in Section~\ref{subsec:GWalgoGauss} is too lengthy, we demonstrate in the following that the tree-center integrals integrals~$(\mu\nu |P)_m$ and the metric matrix~$\mathbf{M}$ originate indeed from the RI factorization of the 4c-CIs introduced in Eq.~\eqref{ERI4c}. 
We will rationalize that the 4c-CIs can be fully recovered with the consequence that our algorithm is exact in the limit of a complete RI basis set. 
\par
For the exchange part of the self-energy, the 4c-CIs can be directly obtained by inserting Eq.~\eqref{DV} into Eq.~\eqref{Sx} and using Eq.~\eqref{ERI4c}, which yields the familiar expression for the exchange self-energy,\cite{GWCP2K}
$\Sigma^\text{x}_n=-\sum_{i}^\text{occ}(ni|in)_\text{RI}$.\par
%
%
%
%
%

%
%
The RI factorization of the 4c-CIs is less obvious for the correlation part $\Sigma_n^c$ of the self-energy and the intermediate steps. We exemplarily show for the matrix elements $W_{PQ}^c$, where the 4c-CIs occur. To this end, we  
%
%
use the Taylor expansion $(1\mt x)^{-1}\eqt 1\pt  x\pt x^2\pt{\ldots}$ to express the inverse of the dielectric function from Eq.~\eqref{dielfunc} as
\begin{align}
    \boldsymbol{\epsilon}^{-1}(i\omega) = \mathbf{1}+\bLT\bchitil(i\omega)\bL+ (\bLT\bchitil(i\omega)\bL)^2+
    \ldots\,\;.
\end{align}
We can then rewrite Eq.~\eqref{Wc} as
\begin{align}
\mathbf{W}^\text{c}(i\omega) 
=& \;\bL\Big[\bLT\bchitil(i\omega)\bL+ (\bLT\bchitil(i\omega)\bL)^2+{\ldots}\Big]\bLT\,\;.
\label{eq:Wtaylor}
\end{align}
After inserting Eqs.~\eqref{chipseudoimtime} and \eqref{VLL} into Eq.~\eqref{eq:Wtaylor}, we obtain
\begin{align}
 \mathbf{W}^\text{c}(i\omega) 
=&\; \bV \chidetails \bV \nonumber
\\ & + \bV\chidetails \bV \chidetails\bV + \ldots
\,.
\end{align}
With Eq.~\eqref{NsquareRPA}, we recover the RI expression~\eqref{ERI4c} in the quadratic term:
\begin{align}
&[\boldsymbol{\chi}^0(i\omega)\mathbf{M}^{-1}\bV
\mathbf{M}^{-1}\boldsymbol{\chi}^0(i\omega)]_{PQ}\nonumber
\\
&
=
-\hspace{-0.3em}
\sum_{RSTU}
\sum\limits_{\mu\sigma\lambda\nu}
\sum\limits_{\overline{\mu\sigma\lambda\nu}}
\text{FT}[G_{\mu\lambda}(i\tau)G_{\nu\sigma}(-i\tau)](i\omega)\;
(\lambda\sigma| P)_m  
(\mu\nu |R)_m
\nonumber\\
&\hspace{0.5em}
\times M^{-1}_{RS}V_{ST}M^{-1}_{TU}\;
\text{FT}[G_{\overline{\mu\lambda}}(i\tau)G_{\overline{\nu\sigma}}(-i\tau)](i\omega)\;
(\overline{\lambda\sigma}| U)_m  
(\overline{\mu\nu} |Q)_m
\\[0.5em]
&
=
-\hspace{-0.3em}
\sum\limits_{\mu\sigma\lambda\nu}
\sum\limits_{\overline{\mu\sigma\lambda\nu}}
\text{FT}[G_{\mu\lambda}(i\tau)G_{\nu\sigma}(-i\tau)](i\omega)\;
(\lambda\sigma| P)_m 
\nonumber\\
&\hspace{0.5em}
\times 
(\mu\nu |\overline{\lambda\sigma})_\text{RI}\;
\text{FT}[G_{\overline{\mu\lambda}}(i\tau)G_{\overline{\nu\sigma}}(-i\tau)](i\omega)\;
(\overline{\mu\nu} |Q)_m\,.
\end{align}

The RI expression~\eqref{ERI4c} can be found in similar fashion for all higher orders in $W^c_{PQ}$ and ultimately also for the expression of the self-energy in Eq.~\eqref{corrSEtaurewrite}. 
%
%
%

\section{Implementation details}
\label{sec:impldetails}
%
%
%
%
%
%
We have implemented the low-scaling $GW$ algorithm outlined in Section~\ref{subsec:GWalgoGauss} in the open-source software package CP2K\cite{Kuehne2020} which is available from github.\cite{cp2kgithub}
%
%
The parallelization of the algorithm is mostly based on the standard message passing interface (MPI). OpenMP threading in a hybrid MPI+OpenMP approach is also supported. All steps of the algorithm have been optimized for massively parallel executation on more than 10,000 CPU cores. Most optimization efforts were dedicated to the computationally most expensive step, the calculation of $\chi_{PQ}^0(i\omega)$, using the concepts outlined in Ref.~\citenum{cubicRPAcp2k} and the DBCSR library for sparse matrix-tensor operations.\cite{dbcsr} DBCSR is also employed for sparse matrix-matrix operations in Eqs.~\eqref{corrSEtaurewrite}, and~\eqref{Sx}. \par
The proper choice and optimization of the imaginary-time and imaginary-frequency grids is crucial for computational efficiency and accuracy. We employ the mini\-max time $\{\tau_j\}_{j=1}^{\Nminimax}$ and frequency  $\{\omega_k\}_{k=1}^{\Nminimax}$ grids with $\Nminimax$ grid points as pioneered by Kaltak~\textit{et al.}\cite{RPAKresse} and Liu~\textit{et al.}\cite{liu2016cubic}
For mini\-max, the a-priori known analytical structure of $\chi$, $W$ and $\Sigma$ is used to construct grids that minimize the $L^\infty$ norm of the error between exact integration and numerical integration. 
Following this procedure, optimal grids can be constructed for the Fourier transforms\cite{liu2016cubic,RPAKresse} of the respective functions~$f$,
\begin{align}
  f(i\omega_k)  &= \sum_{j=1}^{\Nminimax} 
  \gamma_{kj}\sd\exp(i\omega_k\tau_j)\,f(i\tau_j)\,,
  \\
    f(i\tau_j)  &= \sum_{k=1}^{\Nminimax} 
  \xi_{jk}\exp(i\tau_j\omega_k)\,f(i\omega_k)\,.
\end{align}
For simplicity, we compute the weights~$\gamma_{kj}$ and $\xi_{jk}$ during the program execution from $L^2$ minimization.\cite{RPAKresse}
Mini\-max grids are constructed by the Remez algorithm, which requires higher numerical precision than the standard double precision used in electronic-structure calculations. The mini\-max grids are therefore not optimized during run-time, but computed with quadruple precision and pretabulated.\cite{minimaxExp}
%
%
For details on generating mini\-max grids, we refer to the comprehensive literature.\cite{RPAKresse, liu2016cubic} Note that mini\-max grids were recently also developed for  finite-temperature $GW$.\cite{Kaltak2020}\par
In our previous work,\cite{cubicGW} we employed 12 mini\-max points. 
To achieve higher accuracy, we have now computed mini\-max grids with 26, 28, 30, 32, and 34 points in imaginary time and imaginary frequency for different ranges.\cite{RPAKresse} 
These grids are freely available on github~\cite{cp2kgithub} for usage with other codes implementing the space-time method. 
As we demonstrate in Section~\ref{sec:accuracy}, benchmark accuracy is already obtained with 30 mini\-max points. Since the convergence of the Remez algorithm is increasingly difficult with the number of points, the generation of grids with more than 34 points has not been attempted.

\section{Computational details}
\label{sec:compdetails}

%
%
The low-scaling $GW$ calculations are performed with the program package CP2K~\cite{Kuehne2020} and reference calculations are carried out with the program package FHI-aims.\cite{Blum2009} The input and output files of these calculations are available from the Novel Materials
Discovery (NOMAD) repository.\cite{nomad_repo}

\subsection{Low-scaling \textit{GW} calculations using CP2K}
\label{sec:computationaldetailscp2k}
We perform $G_0W_0$ calculations with the low-scaling algorithm on the $GW100$ benchmark set (Section~\ref
{sec:accuracy}) and $G_0W_0$ as well as ev$GW_0$ calculations on phosphorene nanosheets (Section~\ref{sec:alllevels}\,--\,\ref{sec:computeff}). All $GW$ calculations start from all-electron DFT calculations using the Gaussian and augmented plane-waves  scheme (GAPW)\cite{GAPW} and the Perdew-Burke-Ernzerhof (PBE)\cite{PBE} exchange-correlation functional. %
%
%
We use the RI with the truncated Coulomb metric with a truncation radius of $\rc \eqt3$\,\AA, unless otherwise noted.
The self-energy is analytically continued from the imaginary to the real-frequency domain using a Pad\'{e} model\cite{Vidberg1977,GW100,liu2016cubic} with 16 parameters.\par
For the $GW100$ benchmark calculations, we use the def2-QZVP\cite{Weigend2003} basis set as primary basis set and def2-TZVPPD-RIFIT\cite{Haettig2005} as auxiliary basis set. We employ mini\-max grids with $N\eqt30$ time and frequency mini\-max points for the $GW100$ study, unless otherwise stated.\par
%
%
%
%
%
\par
The molecular geometries of the phosphorene nanosheets are obtained as follows: 
We relax the unit cell of free-standing phosphorene using PBE-D3,\cite{PBE,Grimme2010} Goedecker-Teter-Hutter pseudopotentials~\cite{GTH} and a TZVP-MOLOPT basis set\cite{Vandevondele2007} using an 8\,$\times$\,6 $k$-point mesh.
Then, an $L\timest L$ ($L\intext \mathbb{N}$) supercell is formed, periodic boundary conditions are removed and dangling bonds are saturated by hydrogen atoms.
The hydrogen atoms are relaxed with PBE-D3 while keeping the phosphorus atoms fixed. \par
For the $GW$ calculations on phosphorene nanosheets, we employ the all-electron aug-cc-pVDZ basis sets\cite{Dunning1989,Dunning1994,Kendall1992} in combination with the RI basis set aug-cc-pVDZ-RIFIT. \cite{Haettig2005,Weigend2002,EMSL} The lowest exponents of the RI basis set have been scaled up for the calculations on the large phosphorene sheets reported in Section~\ref{sec:appl} to improve the performance; see SI for details. Mini\-max grids with 30 time and frequency points are used for the small phosphorene clusters studied in Section~~\ref{sec:alllevels}, while 14 mini\-max points are used for the large phosphorene sheets. 
In the sparse matrix-tensor operations from Eq.~\eqref{NsquareRPA}, we filter atomic tensor blocks conservatively with a Frobenius norm of the atomic blocks of $10^{-15}$.
For ev$GW_0$, we employ 80 occupied and 80 unoccupied $GW$ levels in the self-consistency loop. %
For levels outside this range,  a constant shift in the ev$GW_0$ cycle has been applied.\par
With these settings, we find that the $G_0W_0$ and ev$GW_0$ HOMO-LUMO gap of the large phosphorene sheets (Section~\ref{sec:appl}) is converged within 0.02\,eV compared to calculations using a fully converged mini\-max grid of 30 points and the aug-cc-pVQZ\cite{Dunning1989,Kendall1992} basis set, see SI for more details. An extrapolation to the complete basis set limit, as often necessary in $GW$, is therefore not required. HOMO-LUMO gaps typically converge faster with respect to basis set size than ionization potentials and affinities, which was demonstrated in, e.g., Ref.~\citenum{Foerster2020a} for subsets of medium and large molecules from the $GW5000$ database.\cite{Stuke2020}
Additionally, we compute the PBE gap of 2D periodic phosphorene from GAPW all-electron calculations using the aug-cc-pVTZ basis set\cite{Dunning1989,Dunning1994,Kendall1992} and an 8\,$\times$\,6  $k$-point mesh.

\subsection{Reference \textit{GW} calculations with contour deformation using FHI-aims}
\label{sec:computationaldetailsfhiaims}
We perform reference DFT calculations with the PBE functional for all phosphorene nanosheets and $G_0W_0$@PBE calculations for the smaller phosphorene nanosheets up to 180 atoms using the FHI-aims program package.\cite{Blum2009} FHI-aims is a native all-electron code based on numeric atomic-centered orbitals (NAOs). For direct comparison with the low-scaling calculations, we employ also the aug-cc-pVDZ Gaussian basis sets, which can be considered as a special case of an NAO and which are treated numerically in FHI-aims. The auxiliary basis sets are constructed ``on-the-fly" by forming product pairs of primary basis functions and subsequent removal of linear dependencies as described in Ref.~\citenum{Ren2012}.\par
The $GW$ calculations are performed with the contour deformation implementation\cite{Dorotheacontroudef} in FHI-aims, unless otherwise noted. As for the low-scaling CP2K calculations, the QP equations are always solved iteratively. In addition to computing the QP energies for the phosphorene nanosheets, we also compute the self-energy matrix elements for a small phosphorene cluster with 24 atoms comparing contour deformation and analytic continuation.\cite{Ren2012} For the latter, we use  the Pad\'{e} approximation with 16 parameters, as in the CP2K calculations. Both methods, contour deformation and analytic continuation, require the computation of integrals over the imaginary frequency axis, for which we employ a modified Gauss-Legendre grid\cite{Ren2012} with 200 grid points. For the Pad\'{e} model, the same set of grid points $\{i \omega\}$ is used to calculate $\Sigma^c_n(i\omega)$.\par
Using the same basis set, the DFT-PBE gaps of the phosphorene sheets agree within 1~meV between CP2K and FHI-aims and the $G_0W_0$ gaps within 20~meV; see Table~II (SI). 

\section{\textit{GW}100 benchmark: accuracy of frontier orbitals}\label{sec:accuracy}
In the following, we assess the accuracy of the low-scaling $GW$ algorithm  for HOMO and LUMO QP energies of molecules from the $GW100$ benchmark set.\cite{GW100}
We carefully study their convergence with the mini\-max integration grid size, the RI basis set size and the truncation radius used for the RI-tC metric.

\subsection{Data set and reference values}
The $GW100$ benchmark set contains HOMO and LUMO energies of 100 small molecules featuring a variety of elements from the periodic table. We exclude the multi-solution cases BN, BeO, MgO, O$_3$ and CuCN from computing the MAD of the HOMOs for the following reasons. First, the real self-energy matrix elements of these molecules exhibit poles in the frequency region of the quasiparticle, leading to at least two different solutions with similar spectral weight.\cite{GW100} Different codes might find equally valid solutions and one should rather compare the self-energy matrix elements, as done in Ref.~\citenum{GW100}. Second, 128 Pad\'{e} parameters are necessary to resolve these poles.\cite{GW100} This implies that $\Sigma(i\omega)$ must be computed on a frequency grid of at least 128 points, which is far beyond the size of currently available mini\-max grids. All 100 molecules are included for the MAD of the LUMO.\par
We use the $G_0W_0$@PBE results from FHI-aims reported in the original $GW100$ work\cite{GW100} as reference. The FHI-aims results from Ref.~\citenum{GW100} were computed with analytical continuation using the Pad\'{e} model approximation with 16 parameters, as in our approach. The analytic-continuation results from FHI-aims are of high numerical quality for frontier orbitals, matching the results from a fully analytic evaluation of the self-energy within a few meV, as shown for a $GW100$ subset in Ref.~\citenum{GW100}. Our goal is to assess the numerical accuracy of the algorithm for a given primary basis set. We therefore compare the data directly at the def2-QZVP level instead of basis-set extrapolated results.
%

%
\subsection{Convergence of mini\-max grid}\label{sec:minimaxcheck}
\begin{table}[t!]
 \fontsize{9}{11}\selectfont
\caption{
Convergence of HOMO and LUMO energies of the $GW100$ benchmark set computed with the low-scaling algorithm at the  $G_0W_0$@PBE level as function of the number of mini\-max points $N$.
%
%
%
Listed are the mean absolute deviations (MADs) with respect to the FHI-aims reference values from Ref.~\citenum{GW100} (16-parameter Pad\'{e} model, def2-QZVP) and the number of excitations (out of 95 for the HOMO and out of 100 for LUMO) with MADs $\le$\,0.01~eV and $\le$\,0.02~eV. 
%
}
\begin{tabular}{@{\extracolsep{\fill}}c@{\hspace{1.0\tabcolsep}}c@{\hspace{1.0\tabcolsep}}c@{\hspace{1.0\tabcolsep}}c@{\hspace{1.0\tabcolsep}}c@{\hspace{1.0\tabcolsep}}c@{\hspace{1.0\tabcolsep}}c}
\hline\hline
\\[-0.8em]
 $\Nminimax$ 
 &\multicolumn{2}{c}{MAD (eV)}
 &\multicolumn{2}{c}{MAD $\le$ 0.01 eV }
 &\multicolumn{2}{c}{MAD $\le$ 0.02 eV }
\\[0.3em]\cmidrule(l{0.05em}r{0.5em}){2-3} \cmidrule(l{0.05em}r{0.5em}){4-5}\cmidrule(l{0.05em}r{0.5em}){6-7}
&HOMOs & LUMOs&HOMOs & LUMOs&HOMOs & LUMOs
\\[0.5em] 
\hline
\\[-0.8em]
 10  & 0.098  & 0.046  & 24 & 26 & 36 & 41
\\[0.2em]
 20  & 0.025   &  0.013  & 32 & 77 & 66 & 96
\\[0.2em]
 26  & 0.014     &  0.009  & 75 & 92 & 84 & 98
\\[0.2em]
 28  &  0.009     &  0.007   & 81  & 94 &  89 & 97 
\\[0.2em]
 30  & 0.007  & 0.006  & 87 & 93 & 92 & 98
\\[0.2em]
 32  & 0.007  & 0.005  & 88 & 94 & 91 & 99
\\[0.2em]
 34  & 0.007  & 0.005  & 92 & 95 &  93 & 98 
\\[0.2em]
\hline\hline
 \end{tabular}
\label{accuracytable}
\end{table}

The convergence of the $G_0W_0$@PBE QP energies with respect to to the size of our generated mini\-max grids is reported in Table~\ref{accuracytable}. Except for different mini\-max parameters, the settings given in Section~\ref{sec:computationaldetailscp2k} were used. The MADs with respect to the $GW100$ reference results decrease quickly with the grid size.
%
%
Already for 28 mini\-max points, we observe an MAD of $<$ 10 meV for both, HOMOs and LUMOs.
The accuracy saturates at 30 mini\-max points with an MAD of 7~meV for HOMOs and 6~meV for LUMOs. The gain of accuracy when employing even larger grids with 32 and 34 points is marginal. Therefore, we set the mini\-max grid with 30 point as default for benchmark studies with the low-scaling algorithm.

The low-scaling $GW$ algorithm reported in Ref.~\citenum{liu2016cubic}, which is the PAW variant of the space-time method, reaches high accuracy already for smaller mini\-max grids. Liu \textit{et al.}\cite{liu2016cubic} showed that 20 time and frequency points were sufficient to reach convergence within 10~meV. As shown in Table~\ref{accuracytable}, the MAD is still larger than 20~meV with the same grid size in our scheme. The different convergence behaviour is probably due to the different treatment of the core electrons. The low-scaling PAW-$GW$ schemes does not treat the core electrons explicitly, which reduces the mini\-max range\cite{RPAKresse} compared to our all-electron scheme. With smaller mini\-max ranges less grid points are generally needed to obtain the same accuracy.
%
%
%

%
%
%

\subsection{Convergence of RI basis sets and truncation radius}\label{sec:tcRIcheck}
\begin{figure}[t!]
\centering
\includegraphics[width=0.48\textwidth]{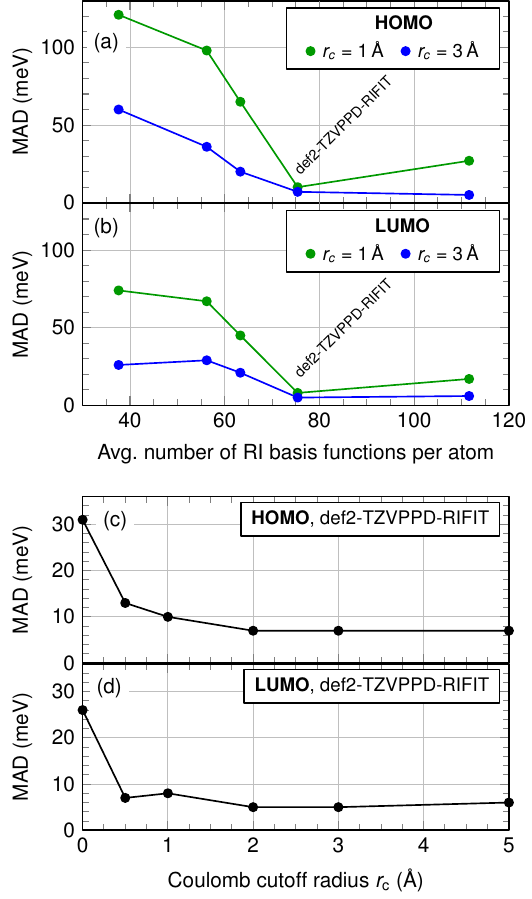}
\caption{
Convergence of HOMO and LUMO energies of the $GW100$ benchmark set computed with the low-scaling algorithm at the  $G_0W_0$@PBE level with respect to (a,b) the RI basis set size and (c,d) the truncation radius from Eq.~\eqref{tC}. Presented are mean absolute deviations (MADs) with respect to the FHI-aims reference data\cite{GW100} (16-parameter Pad\'{e} model, def2-QZVP). For (a) and (b), we employ the RI basis sets def2-SVP-RIFIT, def2-TZVP-RIFIT, def2-TZVPP-RIFIT, def2-TZVPPD-RIFIT and def2-QZVPP-RIFIT.\cite{Haettig2005}
%
%
In (c) and (d), the def2-TZVPPD-RIFIT basis is used as RI basis set. 
}
\label{convRItrunc}
\end{figure}
%
The other two parameters, which influence the accuracy of the low-scaling algorithm, are the RI basis set size and the Coulomb cutoff radius for the RI-tC metric. Both parameters are in principle interdependent since the cutoff radius controls if the metric is more ``overlap-like" or rather resembles the conventional Coulomb metric, which requires smaller RI basis set sizes as discussed in Section~\ref{sec:truncatedcoulomb}.

Figures~\ref{convRItrunc}\,(a) and~(b) show the MAD for the $GW100$ reference data as function of the RI basis set size for HOMO and LUMO, respectively. 
We study the RI basis set convergence for two cutoff values, $\rc \eqt 1\,${\AA} and $\rc \eqt 3\,${\AA}. 
We observe a more consistent convergence behaviour when using the larger cutoff $\rc \eqt 3\,${\AA}.
For $\rc \eqt 1\,${\AA}, the smallest MAD (10~meV) is obtained with the def2-TZVPPD-RIFIT basis. The accuracy becomes worse for larger RI basis sets which might be related to ill-conditioning problems. 
The truncation at $\rc \eqt 3\,${\AA} yields higher accuracy than $\rc \eqt 1\,${\AA} for all RI basis sets that have been tested. The MAD is well below 10~meV for def2-TZVPPD-RIFIT and the next larger RI basis set.

In Fig.~\ref{convRItrunc}\,(c) and~(d), we employ def2-TZVPPD-RIFIT as RI basis and vary the Coulomb cutoff radius used in RI-tC. 
For $\rc \eqt 0\,${\AA}, the RI metric is equivalent to the overlap metric and we obtain an MAD of $\sim$\,30\,meV for the HOMO, which is close to the 35~meV deviation reported in our previous work\cite{cubicGW} for the low-scaling algorithm with the overlap metric.
The accuracy improves when increasing the Coulomb cutoff radius saturating at radii~$\rc \get 2\,${\AA}. This observation and the rapid convergence of the RI basis set with $\rc \get 3\,${\AA} in Fig.~\ref{convRItrunc}\,(a) and~(b) imply that the attractive features of the conventional Coulomb metric are already largely restored at truncation radii between 2\,-\,3\,{\AA}. We choose $\rc = 3\,${\AA} as safe setting for our low-scaling calculations. 
%

%
%
\begin{figure}[t!]
\centering
\includegraphics[width=0.48\textwidth]{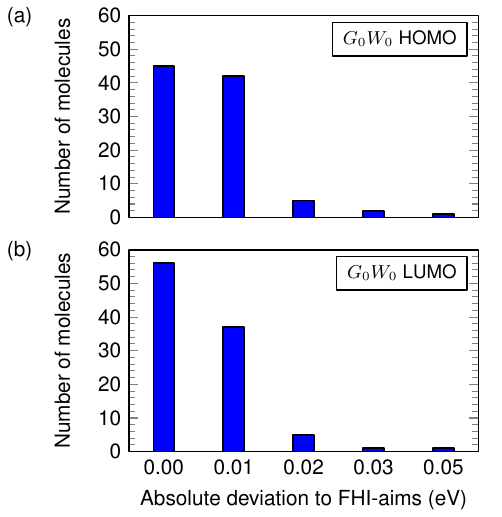}
\caption{
$GW100$ benchmark of the $G_0W_0$@PBE energies computed with the low-scaling algorithm and the settings given in Section~\ref{sec:computationaldetailscp2k} for (a) HOMOs  and (b) LUMOs. Shown are the number of molecules with a given absolute deviation to the FHI-aims values from Ref.~\citenum{GW100} (16-parameter Pad\'{e}, def2-QZVP); see SI for raw data.  
%
%
%
%
}
\label{histogramGW100}
\end{figure}
\subsection{Benchmark with converged settings}
We now compare the $GW100$ results obtained with the settings given in Section~\ref{sec:computationaldetailscp2k}, i.e., the converged settings (30 mini\-max points, def2-TZVPPD-RIFIT, $\rc = 3\,${\AA}) to the FHI-aims reference data. The number of molecules with a given absolute deviation from the reference data are shown in Fig.~\ref{histogramGW100} (see Table~I (SI) for the raw data). We find that 87 out of 95 HOMO energies and 93 out of 100 LUMO energies agree with the FHI-aims reference\cite{GW100} within 10~meV, see also Table~\ref{accuracytable}. Only three excitations for the HOMO and two excitations for the LUMO differ by more than 20~meV from the reference and the maximum deviation is 50~meV.\par
Compared to our old implementation with the overlap metric,\cite{cubicGW} the accuracy is significantly improved. The MAD is reduced from 35~meV to 7~meV for the HOMO and from 27~meV to 6~meV for the LUMO. The MAD is now in the range that was reported for the FHI-aims reference data and the fully-analytic Turbomole results (3~meV for the HOMO and 6~meV for the LUMO).\cite{GW100} 
The RI-tC scheme and the new mini\-max grids also improve the reliability of the low-scaling algorithm. The number of outliers is reduced to zero. In our previous work,\cite{cubicGW} we observed 9 energies with deviations $\get$60~meV for the HOMO and 7 for the LUMO, including a couple of extreme cases with errors of 0.7 and 2.2~eV. 
\section{Accuracy for semi-core and unbound states}\label{sec:alllevels}
\begin{figure}[t]
\centering
\includegraphics[width=0.465\textwidth]{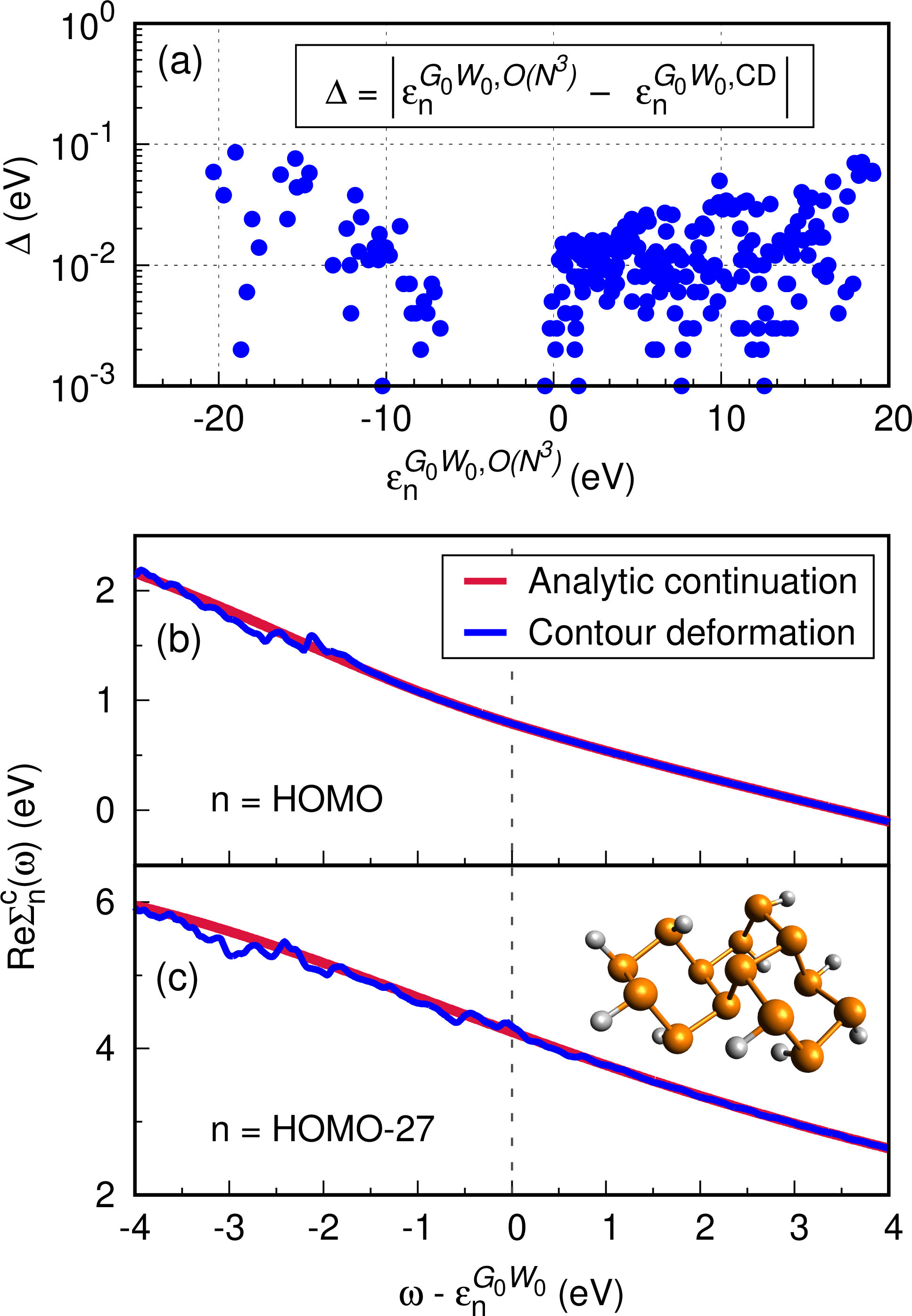}
\caption{
$G_0W_0$ quasiparticle energies of the phosphorene nanosheet H$_{10}$P$_{14}$ including all states between --\,20\,eV to 20\,eV. (a) Absolute deviation $\Delta\,{\coloneqq}\,|\varepsilon_n^{G_0W_0, O(N^3)}- \varepsilon_n^{G_0W_0, \text{CD}}|$ of the $G_0W_0$ energies computed with the low-scaling ($O(N^3)$) algorithm implemented in CP2K with RI-tC (this work) to the contour deformation (CD) implementation in FHI-aims.\cite{Dorotheacontroudef} (b,c) Real part of the correlation self-energy $\Sigma^c(\omega)$ computed with FHI-aims comparing contour deformation and analytic continuation. Diagonal matrix elements $\text{Re}\Sigma^c_n(\omega)=\braket{\psi_n|\text{Re}\Sigma^c(\omega)|\psi_n}$ for (b) the HOMO and (c) the semi-core state HOMO-27. \JW{Note that the ``ripples" in the self-energy in (b) and (c) are broadened, shallow poles.}
}
\label{semicoreandunboundstates}
\end{figure}
The structure of the self-energy matrix elements  $\Sigma^c_n(\omega)$ is typically featureless around the QP solutions for the HOMO and LUMO.\cite{GW100,Dorotheacontroudef,GWreview2019} Achieving benchmark accuracy is thus easier for states close to the Fermi level. A more challenging test for our low-scaling algorithm are semi-core, deep valence and unbound states. In Fig.~\ref{semicoreandunboundstates}\,(a), we report all $G_0W_0$ QP energies within 20\,eV distance to either HOMO or LUMO for a small phosphorene nanosheet cluster (H$_{10}$P$_{14}$). The cluster is shown as inlet in Fig.~\ref{semicoreandunboundstates}\,(c) and its geometry is reported in the SI. The results are compared to QP energies computed with the highly accurate contour deformation technique (CD) implemented in FHI-aims.\cite{Dorotheacontroudef} We have previously shown that the CD technique with the settings described in Section~\ref{sec:computationaldetailsfhiaims} yields without exception the same numerical accuracy as the fully analytic evaluation of the self-energy, including the difficult case of deep core states.\cite{Dorotheacontroudef} By design, the CD techniques is more accurate than the analytic continuation. We set thus the CD results from FHI-aims as reference for our benchmark of semi-core and unbound states.\par


%
%
%
%
We find that all frontier orbitals in the frequency range
HOMO$-$2\,eV and LUMO$+$2\,eV deviate by 
 at most 0.02\,eV, comparing the CD results with the energy value of the low-scaling $GW$ algorithm introduced in this work, see Fig.~\ref{semicoreandunboundstates}\,(a). The deviation increases with increasing distance from the Fermi level. However, the error is for all levels between $-\,20$\,eV and 20\,eV below 0.10\,eV.\par
The increasing deviation is attributed to the analytic continuation technique, which is employed in our low-scaling $GW$ algorithm. In the final step of the algorithm, the self-energy is analytically continued to the real axis by fitting the matrix elements $\Sigma_n^c(i\omega)$ to a multipole model. These models are usually flexible enough to describe frontier orbitals, as shown in Fig.~\ref{semicoreandunboundstates}\,(b). The self-energy is smooth in the frequency region of the HOMO QP energy and $\Sigma_{\textnormal{HOMO}}^c(\omega)$ is perfectly reproduced by the analytic continuation around the HOMO QP energy. For deep valence and semi-core states and unbound states, $\Sigma_n^c$ increasingly acquires features around the QP energy. This is demonstrated for state HOMO-27 of the phosphorene nanosheet cluster in Fig.~\ref{semicoreandunboundstates}\,(c). The real part of $\Sigma_c^n(\omega)$ has shallow poles around $\omega\eqt\varepsilon_n^{G_0W_0}$, which are broadened in our CD calculation. \JW{These broadened poles appear as ``ripples" in the self-energy.} It is practically impossible to reproduce these small pole features with analytic continuation exactly. This is true for canonical $O(N^4)$ as well as low-scaling implementations of the analytic continuation.\par
The acquisition of pole features around the QP energy for states far from the Fermi level is a conceptual problem of $G_0W_0@$PBE. This can be best understood when rewriting the self-energy into its analytic form\cite{FEGW}
\begin{equation}
 \Sigma_n^c(\omega) = \sum_m \sum_{s} \frac{\Braket{\psi_n\psi_m|P_s|\psi_m\psi_n}}{\omega -\varepsilon_m + 
(\Omega_s-i\eta)\mathrm{sgn}(\varepsilon_{\mathrm{F}} -\varepsilon_m)},
    \label{eq:sigma_pole}
\end{equation}
where $m$ runs over all occupied and virtual states and $\eta$ is a broadening parameter. $\Omega_s$ are charge neutral excitations and $P_s$ the corresponding transition amplitudes. 
From Eq.~\eqref{eq:sigma_pole} we directly see that the self-energy~$\Sigma^c_n(\omega)$ has real-valued poles (for $\eta\,{\rightarrow}\,0$) at $\varepsilon_i\mt\Omega_s$ for occupied states and $\varepsilon_a\pt\Omega_s$ for virtual states.
As we discussed in detail in Ref.~\citenum{coreJPCL}, these poles give rise to satellite features, which accompany the QP excitation. The neutral excitations $\Omega_s$, which are close to eigenvalue differences, are underestimated at the PBE level. For occupied states, the PBE orbital energies $\varepsilon_n$ are overestimated and the poles $\varepsilon_i-\Omega_s$ are located at too large (too positive) frequencies and are too close to the QP energy. For virtual states, the reasoning is the same, just with reverse sign, i.e., $\varepsilon_a+\Omega_s$ are at too small frequencies.\par

The problem that $\varepsilon_i-\Omega_s$ are located at too positive frequencies gets progressively worse for deep states, since the difference between the PBE eigenvalues and corresponding QP energies increases in absolute terms. 
%
%
\JW{This behaviour is visible in Fig.~\ref{semicoreandunboundstates}\,(b) and (c) for the calculation with the exact CD technique. The shallow pole structure is for HOMO--27 in the frequency region of the QP energy, $\omega\eqt\varepsilon_n^{G_0W_0}$, whereas for the HOMO the shallow pole structure is located 2\,eV off from the QP energy. 
}
%
%

\begin{figure*}[t]
\centering
\includegraphics[width=\textwidth]{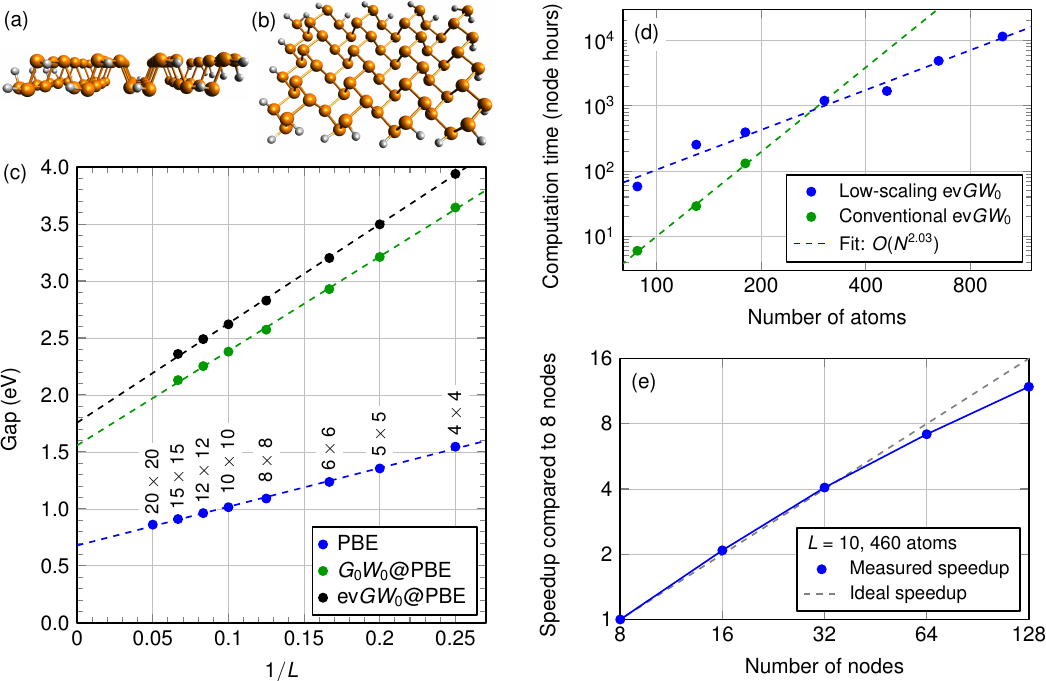}
\caption{
(a) Side and (b) top view of the $4\timest4$ phosphorene nanosheet. 
(c) HOMO-LUMO gap of $L\timest L$ phosphorene nanosheets  computed from DFT-PBE, $G_0W_0$@PBE and ev$GW_0$@PBE as function of the inverse number~$1/L$ of unit cells along an edge of the $L\timest L$ nanosheet.
(d) 
Scaling of ev$GW_0$ execution time with number of atoms for the $L\timest L$ phosphorene nanosheets \JW{comparing the low-scaling implementation from this work to the conventional implementation\cite{GWCP2K} with $O(N^4)$ scaling.} 
Dashed lines are two-parameters least-squares fits of
prefactor and exponent. 
\JW{
%
%
(e) Scaling of the low-scaling $GW$ implementation with respect to number of computing nodes. Presented are strong scaling measurements for the $10\times\,10$ phosphorene sheet (460 atoms) using the cc-pVDZ basis set.\cite{Dunning1989} (Note that aug-cc-pVDZ basis is used in (c) in (d).) 
The calculations in (c), (d) and (e) have been executed on processors of the type "Skylake Xeon Platinum 8174" (48 processors per node) with 96 GB memory [(c) and (d)] and 768 GB memory (e) per node.
%
}
}
\label{ribbons}
\end{figure*}
\par
It has been shown that the correct distance between the poles and the QP solution can be restored in an ev$GW_0$ scheme,\cite{Gatti2015,Zhou2015} even in the extreme case of deep core excitations.\cite{coreJPCL} Since the effect of eigenvalue self-consistency in $G$ is to push the pole structure away from the QP energy,\cite{coreJPCL,GWreview2019} i.e., to more negative and positive frequencies for occupied and unoccupied states, respectively, the self-energy structure is also easier to model by analytic continuation for semi-core states and unbound states. We thus expect that the numerical accuracy of our low-scaling algorithm for non-frontier orbitals is even better than shown in Fig.~\ref{semicoreandunboundstates}\,(a) when using an ev$GW_0$ scheme.
%
%
%


%
\begin{table*}[t]
  \fontsize{10}{12}\selectfont
  \center
  \caption{Fundamental gap of phosphorene in eV calculated from DFT-PBE eigenvalues and $G_0W_0@$PBE and ev$GW_0@$PBE quasiparticle energies. In this work we employ a cluster approach using H-terminated phosphorene sheets consisting of $L\times L$ phosphorene unit cells. The extrapolatd results ($L=\infty$) obtained from Fig.~\ref{ribbons}\,(c) are compared to calculations using periodic phosphorene cells.}
  \begin{tabular*}{0.99\linewidth}{@{\extracolsep{\fill}}cccccc}\toprule
  & \multicolumn{3}{c}{This work: $L\times L$ sheet}  & \multicolumn{2}{c}{Periodic calculation} \\\cmidrule(l{0.5em}r{0.5em}){2-4}  
\cmidrule(l{0.5em}r{0.5em}){5-6}
\centering{method}  & $L=4$ & $L=15$  & $L=\infty$ & This work & Literature 
\\\hline
DFT-PBE          & 1.55 & 0.91 & 0.68 & 0.80 & 0.8 [\!\!\citenum{Frank2019},\citenum{Tran2014a}], 0.90 [\!\!\citenum{Rasmussen2016}]\\
$G_0W_0@$PBE & 3.65 & 2.13 & 1.56 & $-$ & 1.60 [\!\!\citenum{Rudenko2014}], 1.83 [\!\!\citenum{Jiang2017}], 2.0 [\!\!\citenum{Tran2014a}], 2.03 [\!\!\citenum{Rasmussen2016}], 2.06 [\!\!\citenum{Ferreira2017}]\\
ev$GW_0$@PBE & 3.95 & 2.36 & 1.76 & $-$ & 1.94 [\!\!\citenum{Liang2014}], 2.29 [\!\!\citenum{Rasmussen2016}]\\\hline
Experiment &&&&&  2.0 [\!\!\citenum{Liang2014}],  2.2 [\!\!\citenum{Wang2015}]
\\\bottomrule
  \end{tabular*}
  \label{tab:phosporene_gaps}
\end{table*}
\section{HOMO-LUMO gap of phosphorene nanosheets from \textit{GW}}\label{sec:appl}
We apply our low-scaling $GW$ code to finite hyrogen-terminated nanosheets of phosphorene. Phosphorene consists of a single layer of black phosphorus and has been first synthesized in 2014.\cite{Liu2014,Li2014} Phosphorene forms an armchair-like vertically buckled structure of $sp^3$ bonded phosphorus atoms, as shown in Fig.~\ref{ribbons}\,(a) and (b). It has attracted vibrant research interest as two-dimensional semiconductor\cite{Ling2015} because of its direct band gap of $\approx$\,2\,eV at the $\Gamma$ point.\cite{Liang2014,Wang2015} The band gap can be successively decreased from 2\,eV to 0.3\,eV (3D bulk limit) by increasing the number of layers.\cite{CastellanosGomez2015} This band gap range is ideal for many optoelectronic, photovoltaic and photocatalytic applications.\cite{CastellanosGomez2015} \JW{Deformation\cite{stochGWphosphorene} and twisting of layers \cite{Brooks_2020} have been also proposed as methods to modify the band gap of phosphorene.}\par
We show in this work, that the band gap can be also engineered in the  in-plane direction towards values larger than 2\,eV by exploiting finite size effects, which has been recently also reported from Quantum Monte Carlo  calculations.\cite{Frank2019} We study here rectangular hydrogen-terminated phosphorene sheets of size $L\timest L$ $(L\intext\mathbb{N})$, where $L$ indicates the repetition of the phosphorene unit cell, see Fig.~\ref{ribbons}\,(a) and~(b) for a sketch of the molecular geometry. The smallest sheet (4\,$\times$\,4) is of size 1.8\,nm\,$\times$\,1.3\,nm, while the largest (20\,$\times$\,20) is of dimension 9.2\,nm\,$\times$\,6.7\,nm. The progression of the fundamental HOMO-LUMO gaps computed from DFT-PBE eigenvalues and $G_0W_0$@PBE and ev$GW_0$@PBE quasiparticle energies is displayed in Fig.~\ref{ribbons}\,(c). $G_0W_0$ opens the too small PBE gaps, but still suffers from a starting point dependence on the underlying DFT calculation. The $G_0W_0$ gaps are  smaller than the ones from the partially self-consistent ev$GW_0$ scheme, which reduces the dependence on the DFT functional. With all three methods, the computed gaps decrease with increasing sheet size; see also  Table~\ref{tab:phosporene_gaps}. At our highest level of theory, ev$GW_0$@PBE, the HOMO-LUMO gap changes from 3.95~eV (4$\times$4) to 2.36~eV (15$\times$15).
In other words, our calculations indicate that the gap of phosphorene nanosheets can be tuned by more than 1.5~eV when changing the sheet area by a factor of $\sim$\,14.
\par
It is further observed that the PBE, $G_0W_0$ and ev$GW_0$ gaps follow a~$1/L$ behaviour for the $L\timest L$ sheets; see Fig.~\ref{ribbons}\,(c). 
The same $1/L$ scaling has been reported for DFT-PBE computed gaps of 1D-periodic zigzag phosphorene ribbons, whereas an $1/L^2$ has been found for the gaps of their armchair analog.\cite{Tran2014} Our phosphorene sheets feature zigzag as well as armchair edges and we observe here, in agreement with Ref.\citenum{Tran2014}, the dominant scaling of the zigzag edges.\par
In Fig.~\ref{ribbons}\,(c), we extrapolate the gaps towards the 2D bulk limit of phosphorene ($L\rightarrow\infty$). The extrapolated gaps are 0.68 eV (PBE), 1.56 eV ($G_0W_0$) and 1.76 eV (ev$GW_0$). We are confident that our computed gaps of the finite phosphorene sheets are of high numerical quality: In Table II (SI), we show that our gaps are well converged with respect to basis set size. Additionally,  we use a highly accurate full-frequency method for the self-energy evaluation, as we have demonstrated in Section~\ref{sec:alllevels}. However, the comparison of our extrapolated gaps to gaps from periodic $GW$ calculations or the experimentally measured gap of 2D periodic phosphorene (see Table~\ref{tab:phosporene_gaps}) must be taken with a grain of salt. It has been reported in the literature that finite phosphorene sheets host edge states\cite{Peng2014,Liang2014} that are energetically close to the band edges.
These edge states are absent in 2D periodic phosphorene and hence, extrapolating the gap of finite phosphorene sheets may result in a gap that differs from the 2D periodic phosphorene gap. 
As first sanity check, we compare the periodic DFT-PBE gap (0.80\,eV) and the DFT-PBE gap from extrapolation (0.68\,eV) (see Table~\ref{tab:phosporene_gaps}), finding a significant difference of 0.12\,eV. 
We hypothesize that this difference also translates to $GW$ such that our gap extrapolation might underestimate the actual $GW$ 2D bulk limit by at least 0.1~eV.\par
%
%

While our cluster approach might suffer from a conceptual problem for the periodic limit (edge states), periodic $GW$ calculations of 2D systems face several computational challenges as described in Ref.~\citenum{Freysoldt2008} and summarized in the following. Indicative for these numerical challenges is the relatively large spread of the reported periodic $GW$ gaps of 1.6\,-\,2.1\,eV ($G_0W_0$) and  1.9\,-\,2.3\,eV (ev$GW_0$); see Table~\ref{tab:phosporene_gaps}. These variations are most likely due to insufficiencies in the numerical treatment and lack of convergence, which has been systematically studied by Qiu \textit{et al.}\cite{Qiu2016} for a similar system (monolayer of MoS$_2$). For the latter, the reported $GW$ gaps varied within a similar range as for phosphorene. 
\par
One of the computational challenges in 2D-periodic $GW$ calculations is the different screening parallel and perpendicular to the surface, which requires an anisotropic treatment of the singularities of $W$ at the $\Gamma$ point.\cite{Freysoldt2008} A related aspect is that the $k$-point convergence is much slower than for three-dimensional systems, which has been also explicitly shown for phosphorene.\cite{Rasmussen2016} An additional complication is the interaction between the 2D slabs in a 3D periodic approach with plane waves. The vacuum spacing between the repeated slabs cannot be converged out due to the long-range nature of the image charge interaction between the slabs. The correct behavior can be restored by using Coulomb truncation schemes\cite{IsmailBeigi2006,Qiu2016} or post-processing corrections.\cite{Freysoldt2008} All these issues are avoided in our cluster approach, where periodic boundary conditions are not employed.
%
\par

\section{Computational efficiency}
\label{sec:computeff}
\JW{
Finally, we use the phosphorene nanosheets to demonstrate the scaling and the parallel efficiency of our algorithm. As illustrated in Fig.~\ref{ribbons}\,(d), the $O(N^2)$ scaling is preserved from our previous work.\cite{cubicGW} The largest calculations were performed for the phosphorene sheets with 990 atoms (15\,$\times$\sd15 sheet), which corresponds to 6795 electrons per spin that are expanded in 25110 basis functions. 
The crossover between the traditional $O(N^4)$ implementation and the low-scaling $GW$ calculation is at around 300 atoms [$\approx$\,2100 electrons per spin, $\approx$\,7600 basis functions]. As shown in Fig.~\ref{ribbons}\,(d), the crossover point is found by extrapolation due to the high memory demands of the conventional algorithm, which practically restricts the conventional $GW$ calculations to phosphorene sheets of 200\,-\,250 atoms. 
Our low-scaling approach improves also the scaling with respect to memory consumption. The conventional implementation scales $O(N^3)$ in memory, which is reduced to $O(N^2)$ in this work.\par

%

%
In our previous implementation of low-scaling $GW$ with the overlap metric, we reported a crossover point at 150 atoms for quasi-1D graphene nanoribbons.\cite{cubicGW} The shift to 300 atoms is because of the larger amount of three-center integrals, that need to be included in the computation of~$\boldsymbol{\chi}_0(i\tau)$ in Eq.~\eqref{NsquareRPA}. This is caused by three circumstances. First, sparsity conditions are only met for larger system sizes due to the 2D nature of the phosphorene sheets, whereas graphene nanoribbons are quasi-1D systems. Second, we use in this work a Gaussian basis set (aug-cc-pVDZ) with much smaller exponents than in Ref.~\citenum{cubicGW}. The aug-cc-pVDZ basis comprises very diffuse functions (lowest Gaussian exponent in aug-cc-pVDZ for H: 0.02974 a.u., for P: 0.0343 a.u.). For diffuse functions, less three-center integral are zero than for more compact basis sets. Third, the truncated Coulomb metric is "less local" than the overlap metric used in Ref.~\citenum{cubicGW}. All three points increase the computational prefactor, which is the reason, why the largest phosphorene sheet contains ``only" 990 atoms, while in Ref.~\citenum{cubicGW}, we reported $GW$  calculations of a graphene nanoribbon with around 1700 atoms.\par

%
%
%
%
%
%

%
The parallel performance of our low-scaling algorithm is assessed for the 10\,$\times$\sd10 phosphorene sheet (460 atoms) employing the cc-pVDZ basis set.\cite{Dunning1989} Strong scaling measurements for this system are reported in Fig.~\ref{ribbons}~(e), where the speed-up of the calculation with respect to 8 computing nodes is shown. The $GW$ calculation for the 10\,$\times$\sd10 sheet scales well up to 128 nodes (6144 processes) with a parallel efficiency of 74~\%. Note that the $GW$ calculation for the $10\,\times\sd10$ sheet runs also on 2\,--\,7 nodes thanks to an iterative memory reduction scheme.\cite{cubicRPAcp2k} This scheme overcomes memory bottlenecks for small node numbers by additional communication, without increasing the number of operations. Nevertheless, the additional MPI communication slightly increases the computational cost, which  results in a better than ideal speed-up for larger node numbers, which do not require memory reduction. For a fair assessment of the parallel performance, we choose thus 8 nodes as reference in Fig.~\ref{ribbons}~(e) and the cc-pVDZ basis set instead of the aug-cc-pVDZ. The latter is more diffuse and requires more memory than cc-pVDZ, triggering the memory reduction scheme also for node numbers larger than 8.\par
The parallel performance and computational efficiency is also excellent for the larger phosphorene sheets. 
The ev$GW_0$ calculation for the $(15\,\times\sd15)$ sheet (990 atoms) was performed on 768 nodes ($\approx$ 37,000 CPU cores) with a run time of 15\,h.
%
%
%
%
%
%
%
}

\section{Conclusion}
\label{sec:conclusion}
We have presented an accurate low-scaling $GW$ algorithm for computing quasiparticle energies in the $GW$ approximation for systems up to 1000 atoms. 
The algorithm achieves high accuracy by using the RI approach with the truncated Coulomb metric in combination with carefully (pre)optimized mini\-max grids up to 34 time and frequency points each.
We have implemented the method in the open-source quantum chemistry package CP2K\cite{Kuehne2020} and benchmarked the accuracy for HOMOs and LUMOs using the $GW$100 test set. The MADs with respect to the reference values from canonical $GW$ implementations are 7\,meV and 6\,meV, respectively. 
The benchmark studies have been extended to semicore states and unbound unoccupied states using a 24-atom phosphorene cluster. We have shown that
all $GW$ quasiparticle levels in the range between HOMO-20~eV and LUMO+20~eV agree with the highly accurate contour-deformation results from FHI-aims within 0.10\,eV. 
The reported high accuracy together with the good scalability to 1000 atoms is yet another stepping stone towards predictive $GW$ calculations on nanostructured materials. We have demonstrated this on the example of phosphorene, showing that finite size effects can be used to engineer its band gap. 
%

\begin{acknowledgement}
\fontsize{10}{12}\selectfont
We kindly thank Mauro Del Ben, Ferdinand Evers, Jaroslav Fabian, Tobias Frank, Jürg Hutter and Jonas Schramm for helpful discussions. 
The Gauss Centre for Supercomputing is acknowledged for providing computational resources on SuperMUC-NG at the Leibniz Supercomputing Centre under the project IDs pn69mi and pn72pa. We also thank the CSC - IT Center for Science for providing computational resources. 
J.~Wilhelm acknowledges funding from DFG SFB 1277 (project A03).
P.~Seewald acknowledges funding by the NCCR MARVEL, funded by the Swiss National Science Foundation. 
D.~Golze acknowledges financial
support by the Academy of Finland (Grant No.~316168).
\end{acknowledgement}

\begin{suppinfo}
\fontsize{10}{12}\selectfont
In the SI, we show that the RI factorization in a plane-wave RI basis set is independent of the RI metric.
We report the results for the $GW$100 benchmark set with 30 mini\-max points. We provide an input file of CP2K for the $GW$100 test, the xyz geometry of the test of Fig.~3, the customized RI basis set for hydrogen and phosphorus and a CP2K input for a large-scale $GW$ calculation.
Moreover, a detailed comparison between FHI-aims and CP2K on the phosphorene sheets from Section~\ref{sec:appl} is given.
\end{suppinfo}
%


\bibliography{Literature}
\begin{tocentry}
\begin{center}
 \includegraphics[height=3.4cm]{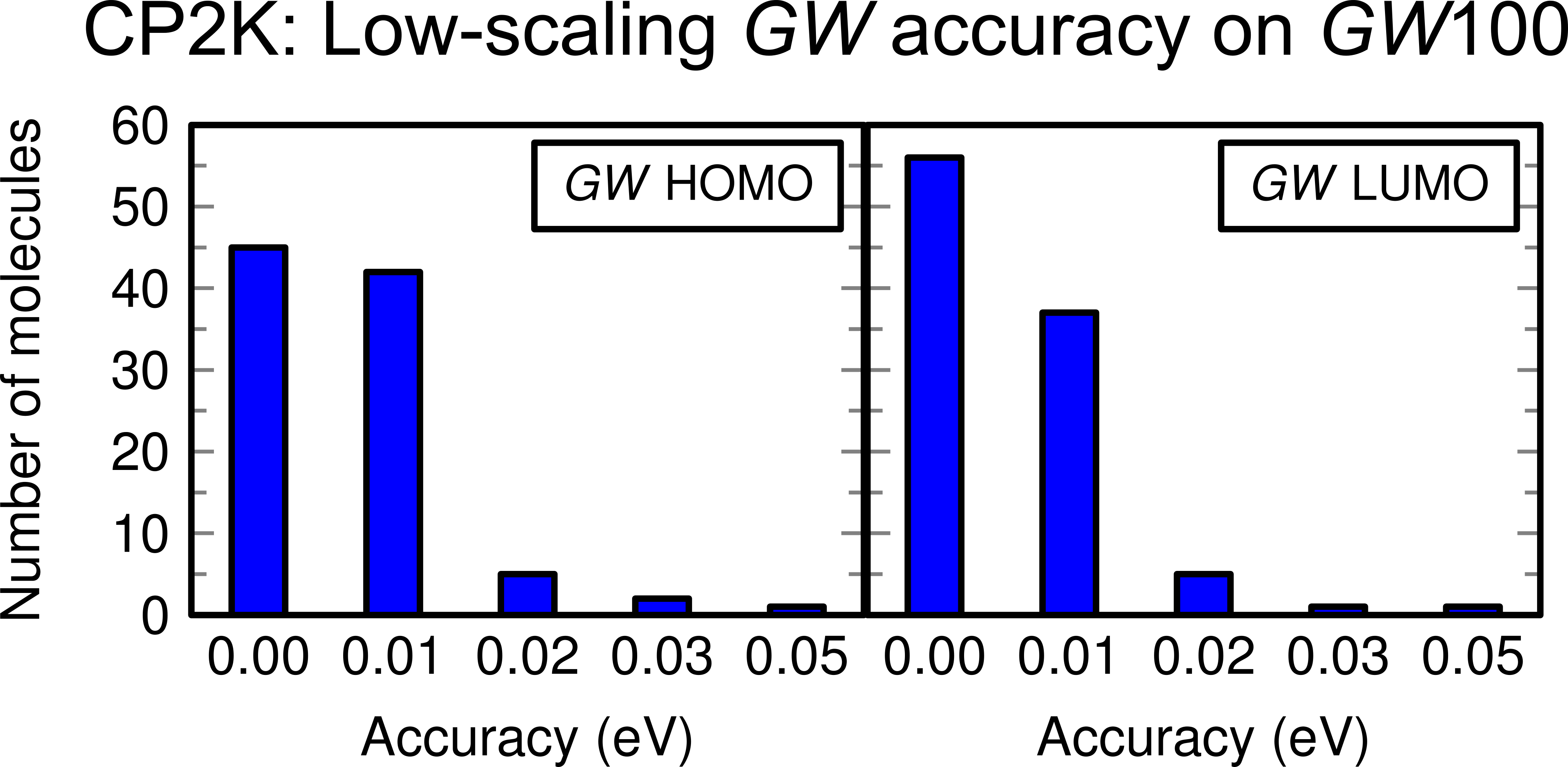}
\end{center}
\end{tocentry}

\end{document}